\documentclass[%
 reprint,
 amsmath,amssymb,
 aps,
 pre
]{revtex4-2}

\usepackage{graphicx}
\usepackage{subcaption}
\usepackage{float}
\usepackage{dcolumn}
\usepackage{bm}
\begin{document}

\preprint{APS/123-QED}

\title{Fine Particle Percolation Dynamics in Porous Media}%

\author{Dhairya R. Vyas$^1$}
\author{Richard M. Lueptow$^{1,2,3}$}
\author{Julio M. Ottino$^{1,2,3}$}
\author{Paul B. Umbanhowar$^1$}
 \email{Contact author: umbanhowar@northwestern.edu}
\affiliation{%
$^1$Department of Mechanical Engineering, Northwestern University, Evanston, Illinois 60208, USA.
\\
$^2$Department of Chemical and Biological Engineering, Northwestern University, Evanston, Illinois 60208, USA.
\\
$^3$Northwestern Institute on Complex Systems (NICO), Northwestern University, Evanston, Illinois 60208, USA.
}%



\begin{abstract}
    The influences of restitution coefficient, $e_n$, inter-particle friction, $\mu$, and size ratio, $R$, on gravity-driven percolation of fine particles through static beds of larger particles in the free-sifting regime ($R \gtrsim  6.5$) remain largely unexplored. Here we use discrete element method simulations to study the fine particle percolation velocity, $v_p$, and velocity fluctuations, $v_{rms}$, for $7 \le R \le 50$ and a range of $e_n$ and $\mu$. Increasing $e_n$ increases velocity fluctuations and reduces percolation velocity. Increasing $\mu$ decreases $v_{rms}$ but its influence on $v_p$ varies with $v_{rms}$, decreasing $v_p$ for low $v_{rms}$ and increasing $v_p$ for high $v_{rms}$. Although the influence of size ratio is  weaker, larger values of $R$ increase both $v_p$ and $v_{rms}$. We also assess the influence of different excitation mechanisms, specifically using static, randomly excited, and sheared beds, finding that an inverse correlation between $v_p$ and $v_{rms}$ persists across all cases and is well-described by the Drude model, where increased scattering reduces mobility, when $v_{rms}$ is large. However, for weakly excited particles with low $v_{rms}$, the Drude analogy breaks down. In this regime, we introduce a staircase-inspired model that accounts for the gravitationally dominated percolation behavior. These findings provide fundamental insight into the mechanisms governing percolation dynamics in porous media and granular systems.
\end{abstract}

\maketitle

\section{Introduction}

Granular materials, comprising particles ranging from micrometers to meters in size, constitute the second most widely used material class in industrial and environmental applications after water~\cite{duranIntroduction2000}. They are central to a wide array of systems, including pharmaceutical manufacturing~\cite{sarkarRoleForcesGoverning2017}, mining and construction~\cite{johnParticleBreakageConstruction2023}, chemical and food processing~\cite{horabikParametersContactModels2016,floreAspectsGranulationChemical2009}, and planetary surface evolution~\cite{meloshMechanicsLargeRock1987,grayRapidGranularAvalanches2003,sanchezSimulatingAsteroidRubble2011}. In many of these applications, the particles are highly polydisperse, giving rise to size-driven segregation that can alter transport, mixing, and mechanical behavior.
A particularly important mode of size-driven segregation involves the downward migration of fine particles through a relatively static bed of larger ones. This mechanism arises in diverse systems, such as powder blending, packed bed reactors, and additive manufacturing, as well as in porous media. In environmental and subsurface flows, the movement of smaller constituents through coarser granular frameworks governs critical processes such as sediment transport in rivers, contaminant dispersion in groundwater, clogging of filtration media, and infiltration in biochar-amended soils~\cite{Fang2024CloggingTO,Sharma1987FinesMI}. Although useful in applications like particle classification or separation, uncontrolled migration of smaller particles can degrade system performance by disrupting flow pathways or inducing structural instabilities~\cite{Zhu2009}.

These dynamics become particularly complex when particle size distributions span several orders of magnitude. For example, fragmented ore in mining ranges from meter-scale blocks to micron-scale fragments~\cite{shinEffectBallSize2013}; soil and sediment beds include both coarse grains and clay-sized particles~\cite{hattoriRockFragmentationParticle1999}; and planetary regolith contains constituents varying in size from nanometers to decimeters~\cite{crostaFragmentationValPola2007}. Similar polydispersity is encountered in advanced manufacturing, where smaller particles percolate through coarser powders during powder bed fusion or composite fabrication, affecting layer uniformity and packing fraction~\cite{shekunovParticleSizeAnalysis2007,tongNumericalStudyEffects2010}. In porous geological media, the infiltration of colloids, silt, or microplastics through granular soils and aquifers can significantly influence contaminant mobility, hydraulic conductivity, and long-term material stability~\cite{Fang2024CloggingTO}.

Many studies have investigated segregation of granular materials, examining the effects of particle size, density, and shape on segregation dynamics in dry systems where the effects of the interstitial fluid (typically air) can be ignored. Models to quantify the segregation forces and velocities have also been developed~\cite{Umbanhowar2019ModelingSI,fan_modelling_2014,gray_particle_2018,deng_modeling_2020}.
Most of the work has focused on systems where the size ratio between large and small particle diameters is relatively small,  $R < 3$. In such systems, segregation is primarily governed by shear in flowing granular materials, which can open interstitial spaces between large particles into which small particles fall and eventually accumulate in lower regions of the flow with large particles accumulating above them.
However, as the size ratio increases, particularly for $R > 6.5$, the segregation dynamics change significantly. In this regime, smaller particles, often referred to as fine particles or simply fines (rather than small particles), percolate through the voids in a static bed of large particles, even with no flow-related shear.  This process, known as free sifting, is a fundamentally different segregation behavior than what occurs in systems with smaller size ratios where flow-related shear is necessary for segregation.

Early experiments investigating free sifting of fine particles were conducted over 50 years ago by Bridgewater and co-workers~\cite{Bridgwater1969,Bridgwater1971}. These studies focus on measuring the percolation velocity, diffusion, and residence time of fine particles percolating through a static bed of large particles. In static beds, free sifting occurs only when the fine particles are smaller than the smallest pore throats, which represent the narrowest constrictions within the interconnected pore network~\cite{gaoPercolationFineParticle2023a}.  While factors such as bed particle polydispersity, particle deformation, and particle shape influence the size of pore throats~\cite{vyasImpactsPackedBed}, in the case of monodisperse spherical bed particles of diameter $D$ with negligible deformation, the smallest pore-throat is formed between three contacting bed particles and allows fine particles of diameter $d$ with size ratio $R=D/d \gtrsim 6.5$ to percolate freely.
Experimental and computational studies  find that fine particles rapidly reach a steady-state percolation velocity~\cite{Bridgwater1969,Bridgwater1971,Rahman2008} that  increases with increasing $R$ ~\cite{Li2010,roozbahaniMechanicalTrappingFine2014a,gaoPercolationFineParticle2023a} and can be related to damping
in a spring-dashpot model using an empirical correlation~\cite{Zhu2009}.
It is also observed that the dispersion of fine particles is diffusive in all directions~\cite{lomineTransportSmallParticles2006,lomineDispersionParticlesSpontaneous2009c}.

Recent investigations examining intermediate size ratios falling between shear-induced percolation and free sifting ($4 \le R \le 7.5$) provide correlations for estimating percolation velocity based on particle size and pore throat size distributions~\cite{gaoPercolationFineParticle2023a} for non-interacting fine particles (i.e., in the single fine particle limit). For a fine particle in a sheared bed with size ratio ranging from $2 \le R \le 10$, two main factors influence the percolation velocity, the opening and closing of pore throats due to shear, and the fine particle velocity fluctuations~\cite{gaoVerticalVelocitySmall2024}. Increased shear rates open interstitial spaces (pores) between large particles  more frequently facilitating percolation, but at high enough shear, excite fine particles, which reduces percolation. 
This behavior, where increased fluctuation leads to reduced percolation, closely parallels electron mobility in metals as described by the Drude model~\cite{Drude}, where increased scattering lowers electrical conductivity. Drawing from this analogy, velocity fluctuations can be treated as a granular counterpart to thermal excitation, with percolation velocity playing the role of drift mobility, thereby capturing the inverse relationship between percolation velocity and velocity fluctuations in sheared granular systems, where both excitation and pore-throat evolution are intertwined~\cite{gaoVerticalVelocitySmall2024}.

To isolate the effect of particle excitation from pore-throat dynamics, we focus on static beds with size ratios $R > 6.464$, where free sifting dominates and pore geometry remains fixed. We vary the normal restitution coefficient, $e_n$, which represents the energy loss due to inelastic deformation during particle collisions, and the friction coefficient, $\mu$, which governs the degree of energy dissipation through sliding interactions between particles, and determine their influence on the fine particle percolation and fluctuation velocities. 
While the findings are relevant to porous media systems such as sediment transport and groundwater contamination, where fine particles migrate through static granular beds, this study focuses on the particle dynamics by excluding the presence of interstitial fluids such as air or water.

We investigate this problem using Discrete Element Method (DEM)~\cite{cundallDiscreteNumericalModel1979b,osullivanParticulateDiscreteElement2014} simulations. 
A primary challenge in DEM simulation is the rapid increase in computational cost as the size ratio increases due to the need to detect potential contacts between neighboring particles~\cite{bergerChallengesIIWide2014}. 
However, computational advances over the past decade have facilitated more efficient simulations that make feasible the study of physically relevant systems with size ratios as high as $R =$ 200~\cite{montiLargescaleFrictionlessJamming2022,montiFractalDimensionsJammed2023,ogarkoFastMultilevelAlgorithm2012}, although a revised velocity-Verlet scheme is necessary to avoid unphysical results at large $R$~\cite{vyasImprovedVelocityVerletAlgorithm2024}. 

In this paper, DEM simulations are used to examine fine particle percolation in the free-sifting regime for size ratios $7 \le R \le 50$. We first show how percolation velocity and velocity fluctuations depend on restitution coefficient, size ratio, and friction coefficient  in static beds of randomly packed large particles. Based on these results, we develop simple scaling relations that capture the effects of dissipation and geometry ($R$). We  demonstrate that the inverse relationship between percolation velocity and velocity fluctuations holds under different excitation mechanisms including static, randomly excited, and sheared beds. The Drude model accurately describes this inverse relationship when velocity fluctuations are large, but not when they are small.  In the latter case,  a staircase-inspired model captures the percolation dynamics in the low-excitation regime.

\section{Simulation setup}

\begin{figure}
        \includegraphics[width=0.8\linewidth]{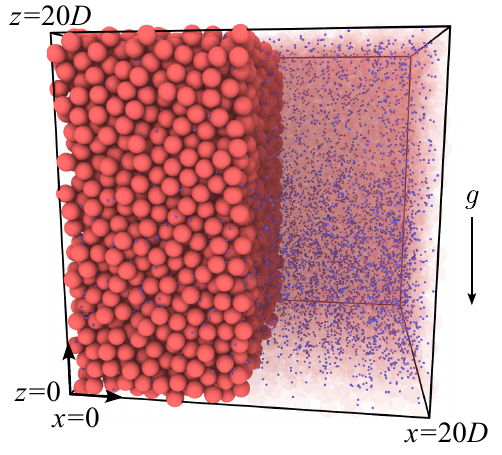}
    \caption{Simulation setup showing $10^4$ fine particles (which only interact with the large particles but not with each other) percolating in a static fully-periodic bed of large particles with large particle diameter $D = 4$~mm and $R = 7$. Bed particles for $x > 10D$, are rendered transparent to show the fine particles. }
    \label{fig:sch}
\end{figure}

The simulation setup is shown in Fig.~\ref{fig:sch}. Large particles, with diameter $D = 4$ mm, are packed in a fully periodic domain of dimensions $20D \times 20D \times 20D$, with a volume packing fraction of 0.60. 
DEM simulations are conducted using the open-source software LAMMPS~\cite{thompsonLAMMPSFlexibleSimulation2022} utilizing the improved velocity-Verlet integration scheme for large size ratios \cite{vyasImprovedVelocityVerletAlgorithm2024}. The \texttt{hertz/material} contact model~\cite{thorntonInvestigationComparativeBehaviour2011c}, coupled with the \texttt{coeff\textunderscore restitution} damping model for  inelasticty~\cite{vyasImprovedVelocityVerletAlgorithm2024} is used to model the normal forces between particles. Tangential friction is modeled using the \texttt{mindlin} model~\cite{Mindlin1949ComplianceOE}, in combination with the \texttt{marshall} twisting model~\cite{MARSHALL20091541}. The material properties, including the Young's modulus and Poisson's ratio, are chosen such that the overlap during contact between large and fine particles remains below 0.1\% of the fine particle diameter for size ratios up to $R = 50$~\cite{vyasImprovedVelocityVerletAlgorithm2024}. This requires a Young's modulus of 70 MPa with a Poisson's ratio of 0.3. To ensure adequate temporal resolution for all size ratios, a timestep of 0.2~\textmu s is used.
For efficient neighbor detection at large $R$, the hierarchical multigrid algorithm in LAMMPS is used \cite{shireSimulationsPolydisperseMedia2021}. 

To generate randomly packed large particle beds, the sizes of randomly distributed bed particles are gradually increased from 0.5 mm to 4 mm within a periodic domain in the absence of gravity~\citep{lubachevskyGeometricPropertiesRandom1990a, gaoPercolationFineParticle2023a}, thereby eliminating the influence of hydrostatic pressure gradients. During this initialization growth phase, a high particle stiffness limits particle overlaps to less than 0.05\% of $D$, ensuring that the pore throat sizes remain unaffected by particle overlap~\citep{vyasImpactsPackedBed}. After the particles reach their target sizes, they are annealed for two seconds and then frozen in place resulting in a large particle packing fraction of 0.60.  This configuration allows for the free percolation of fine particles with $R > 6.5$.

To start a simulation, $10^4$ fine particles are introduced from approximately 1 mm above the top of the static bed under the influence of gravity ($g = 9.81$ m/s$^{2}$) acting in the negative $z$-direction. 
To understand the free-sifting behavior in the single particle limit, fine particles interact only with the large bed particles and not with each other.
Once all the fine particles have entered the bed, the top and bottom periodic boundaries of the simulation box are adjusted to match the dimensions of the bed particle domain rather than just above or just below it, which is necessary to initially accommodate the fine particles entering the domain from just above the bed. These adjusted periodic boundary conditions allow fine particles exiting the bottom  of the bed to repeatedly re-enter the top of the bed. A total of nearly 2200 simulations are performed for $R$ varying from 7 to 10 in increments of 1, and from 10 to 50 in increments of 5.
The restitution coefficient between large and fine particles is varied from 0.1 to 0.99, and the friction coefficient is varied from 0 to 0.3.

The distance required for fine particles to attain steady-state velocity varies with the restitution coefficient: at low $e_n$, steady state occurs after a fine particle travels $2$--$4\,D$, whereas at high $e_n$ it requires up to $20\,D$. In all cases, the steady-state velocity is attained in under one second. 
The mean percolation velocity, \( v_p \), defined as the magnitude of the fine particle's \( z -\)velocity, and the root-mean-square (RMS) fluctuation velocity, $v_{rms}$, which characterizes the excitation of the fine particles, are both calculated using ensemble and time averaging over \(10^4\) fine particles. Temporal averages are taken after the particles reaches a steady-state velocity  ($t > 1$\,s) up to the time that particles are trapped (stop moving downward due to friction~\cite{vyasImprovedVelocityVerletAlgorithm2024} as discussed in Sec.~\ref{sec:friction}) or when simulation ends ($t=5$\,s). 

\section{Results}

\subsection{Restitution coefficient effects}

During steady state percolation the gravitational potential energy gained by the fine particle during its descent is balanced by the energy dissipated through both inelastic collisions and friction. To focus initially on dissipation from inelastic collisions alone, we set $\mu = 0$. This allows us to investigate the effects of varying $e_n$, which modulates  the degree of energy loss during inelastic collisions, and $R$ on the mean percolation velocity $v_p$ and the  fluctuation velocity $v_{rms}$.  We consider the effect of non-zero friction later in Sec.~\ref{sec:friction}.

\begin{figure}[tbh]
    \centering
        \includegraphics[width=\linewidth]{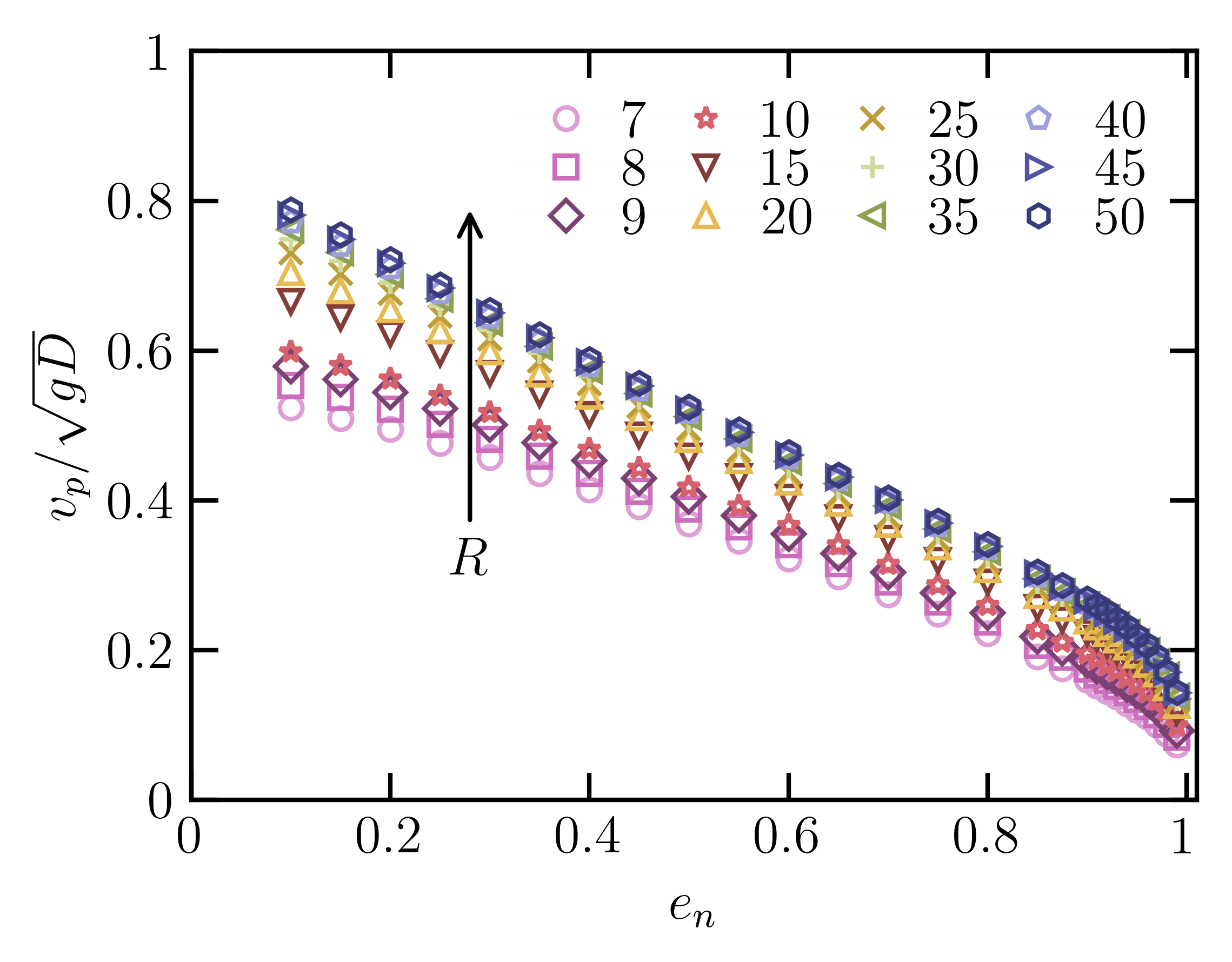}
    \caption{Scaled percolation velocity, $v_p/\sqrt{gD}$, vs\ restitution coefficient, $e_n$, for size ratios $7<R<50$.}
    \label{fig:vp}
\end{figure}

\begin{figure}[tbh]
    \centering
        \includegraphics[width=\linewidth]{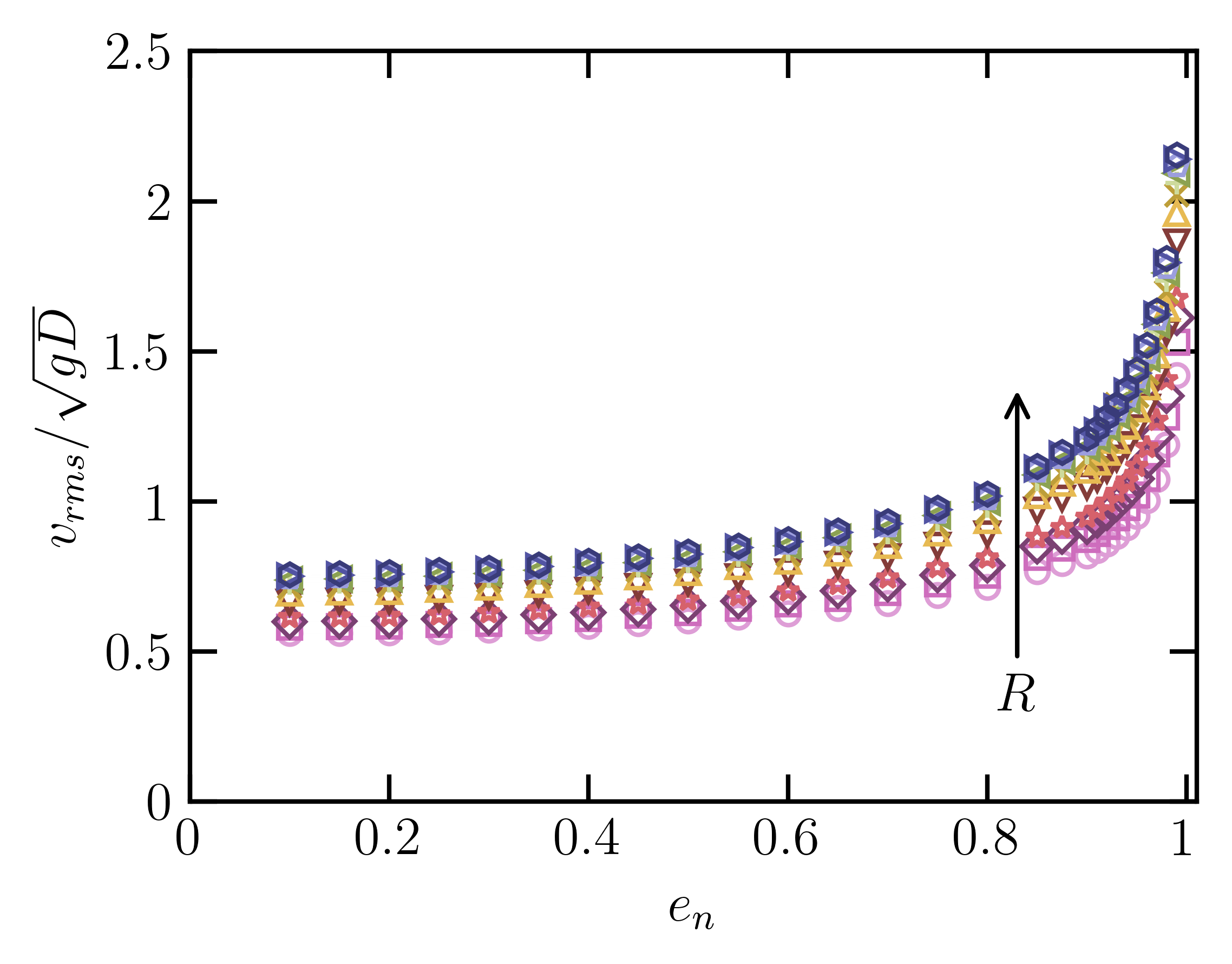}
    \caption{Scaled fluctuation velocity, $v_{rms}/\sqrt{gD}$, vs\ restitution coefficient, $e_n$, for size ratios, $7<R<50$ (symbols as in Fig.~\ref{fig:vp}).}
    \label{fig:vrms}
\end{figure}

The restitution coefficient is varied from 0.1 to 0.85 in increments of 0.05, from 0.85 to 0.9 in increments of 0.025, and from 0.9 to 0.99 in increments of 0.01, for different size ratios. The resulting variation in $v_p$ and $v_{rms}$ with $e_n$ is shown in Fig.~\ref{fig:vp} and Fig.~\ref{fig:vrms}, respectively. 
The velocities are non-dimensionalized using the characteristic velocity scale $\sqrt{gD}$~\cite{Bridgwater1971,gaoPercolationFineParticle2023a}.

Figure~\ref{fig:vp} shows that  $v_p$ decreases toward zero at an increasing rate as $e_n \to 1$. For $e_n \lesssim 0.8$, the decrease is nearly linear, whereas for $e_n \gtrsim 0.8$ the relationship becomes  nonlinear and steeper. A different trend is observed in the fluctuation velocity, $v_{rms}$, in Fig.~\ref{fig:vrms}. As $e_n$ increases, $v_{rms}$ increases. 
This increase is very gradual for $e_n \lesssim 0.8$, but grows sharply once $e_n \gtrsim 0.8$, mirroring the non-linearity of $v_p$. 
These observations suggest the possibility of two different regimes, a low restitution regime ($e_n \lesssim 0.8$) where the change in $v_{rms}$ is minimal and gravity-driven motion dominates, and a high-restitution regime ($e_n \gtrsim 0.8$) where the influence of fluctuations characterized by $v_{rms}$ is more dominant.

\begin{figure}[tbh]
    \centering
        \includegraphics[width=\linewidth]{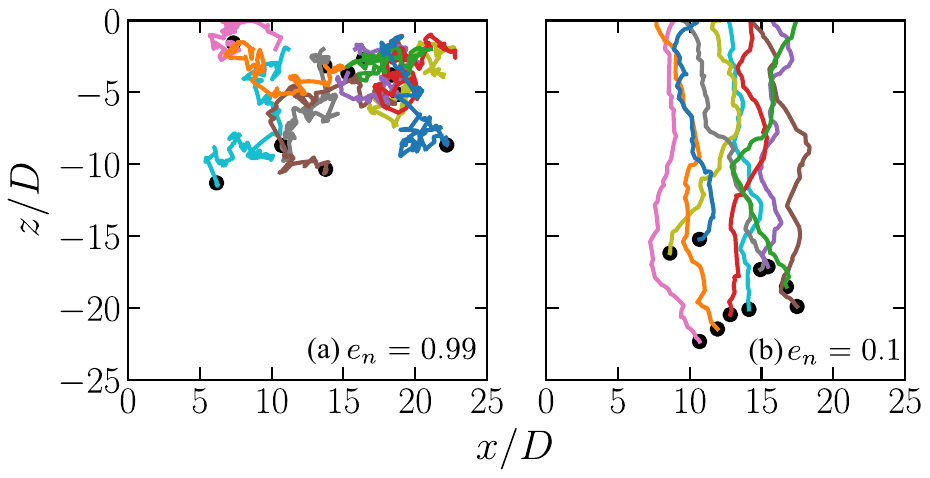}
    \caption{Sample trajectories of ten fine particles with (a) $e_n = 0.99$ (left) and (b) $e_n = 0.1$ (right) for $\mu = 0$ in a static bed of large particles with $R=50$.}
    \label{fig:traj}
\end{figure}

The trajectory plots in Fig.~\ref{fig:traj} for ten fine particles with $R=50$ over 0.5\,s visually demonstrate this  contrast. In the fluctuation-dominated regime ($e_n = 0.99$), $v_{rms}/\sqrt{gD}>1$ and fine particles collide many times with the large bed particles before descending a distance on the order of $D$ [see Fig.~\ref{fig:traj}(a)]. This increased random motion impedes percolation. Consequently, these fines percolate more slowly resulting in a steeper decrease in $v_p$ with $e_n$ in Fig.~\ref{fig:vp} for $e_n \gtrsim 0.8$. In the gravity-dominated regime ($e_n = 0.1$), fines descend more directly [see Fig.~\ref{fig:traj} (b)] and $v_{rms}$ remains more or less constant (as seen in Fig.~\ref{fig:vrms} for $e_n \lesssim 0.8$). In this gravity-dominated regime, restitution controls how the inelasticity of fine particle collisions shapes their bouncing behaviour during repeated gravitational acceleration and subsequent collisions. 
The details of these two mechanisms along with the dashed curves in Figs.~\ref{fig:vp} and \ref{fig:vrms} are discussed in depth in Secs.~\ref{sec:drude} and \ref{sec:staircase}.

\subsection{Size ratio effects}
\label{sec:R}

Even though the system operates in the free-sifting regime and no fine particles are trapped when $\mu = 0$, a size-ratio dependence is also observed in Figs.~\ref{fig:vp} and~\ref{fig:vrms}. In Fig.~\ref{fig:vp}, as $R$ is increased, $v_p$ increases because smaller  fine particles pass through pore throats more easily. The rate of increase of $v_p$ is more pronounced for smaller $R$. For $R>30$, the fine particles are sufficiently small that further reductions in their size minimally affect their percolation behavior, as indicated by the overlapping data points in Fig.~\ref{fig:vp}. 
The fluctuation velocity also increases with increasing $R$, see Fig.~\ref{fig:vrms}. Similar to the results for $v_p$ in Fig.~\ref{fig:vp} the dependence of $v_{rms}$ on $R$ is minimal for $R > 30$.

\begin{figure}
    \centering
        \includegraphics[width=0.8\linewidth]{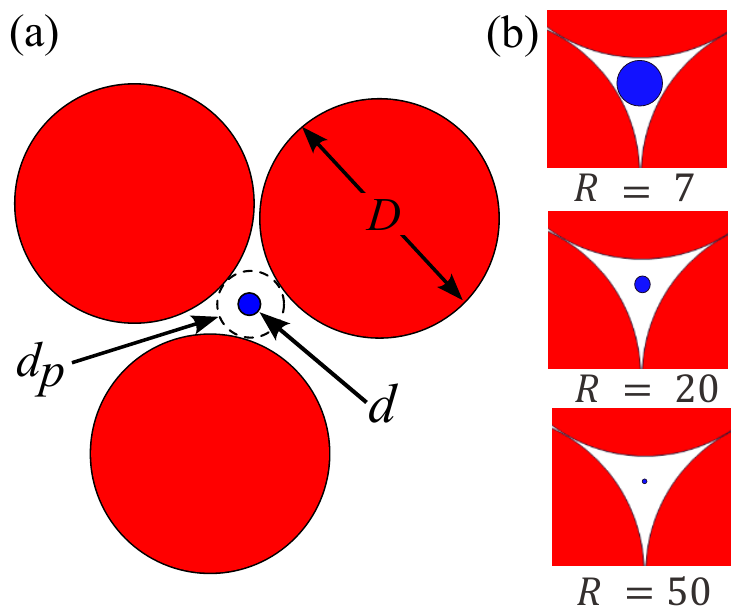}
    \caption{Sketches showing (a) the clearance for a fine particle of diameter $d$ passing through a pore throat of effective diameter $d_p$ formed by a random arrangement of bed particles of diameter $D$, and (b) illustration of the increasing free space with increasing $R$ in a pore throat created between three contacting bed particles.}
    \label{fig:rscale1}
\end{figure}

To develop a size-ratio scaling for $v_p$ and $v_{rms}$, consider how the  size ratio affects the ability of a fine particle to pass through a pore throat between large particles. The schematic in Fig.~\ref{fig:rscale1}(a) shows a general case of a fine particle, with diameter $d$, passing through a pore throat of diameter $d_p$ formed by a random arrangement of bed particles  that may not necessarily be touching one another. The accessible open area of a pore throat for large particles that are in contact with each other depends on the size ratio, as shown in Fig.~\ref{fig:rscale1}(b). A particle with $R = 50$ has significantly more free space to pass through compared to one with $R = 7$, which is just slightly above the free-sifting limit of $R \approx 6.464$.  

To quantify how readily a fine particle passes through a pore throat for bed particles that may not be touching [see Fig.~\ref{fig:rscale1}(a)], we calculate the clearance,  $d_p - d$, where $d_p$ is the pore throat diameter and $d$ is the fine particle diameter.  The ratio of the clearance to the pore throat diameter is
\begin{equation}
    \frac{d_p -d}{d_p} = 1-\frac{R_p}{R},
    \label{eq:clearance}
\end{equation}
where $R_p = D/d_p$ is the pore-throat size ratio and $R=D/d$ is the particle size ratio. 
We assume without justification that the large particle diameter, $D$, used in the velocity scaling, $\sqrt{gD}$, can be replaced by $D(1-R_p/R)$ because of the role that the relative clearance  (i.e., the approximate mean free path in the pore throat) certainly plays in determining both $v_p$ and $v_{rms}$, noting that $\sqrt{gD}$ is recovered for large $R$, as would be expected. Thus, we scale the percolation  and fluctuation velocities as
\begin{equation}
    v_p^* =\frac{v_p}{\sqrt{gD(1 - R_p/R)}} \;\;\text{and}\;\; v_{rms}^* = \frac{v_{rms}}{\sqrt{gD(1 - R_p/R)}}.
    \label{eq:Rscaling}
\end{equation}

\begin{figure}[h]
    \centering
        \includegraphics[width=0.8\linewidth]{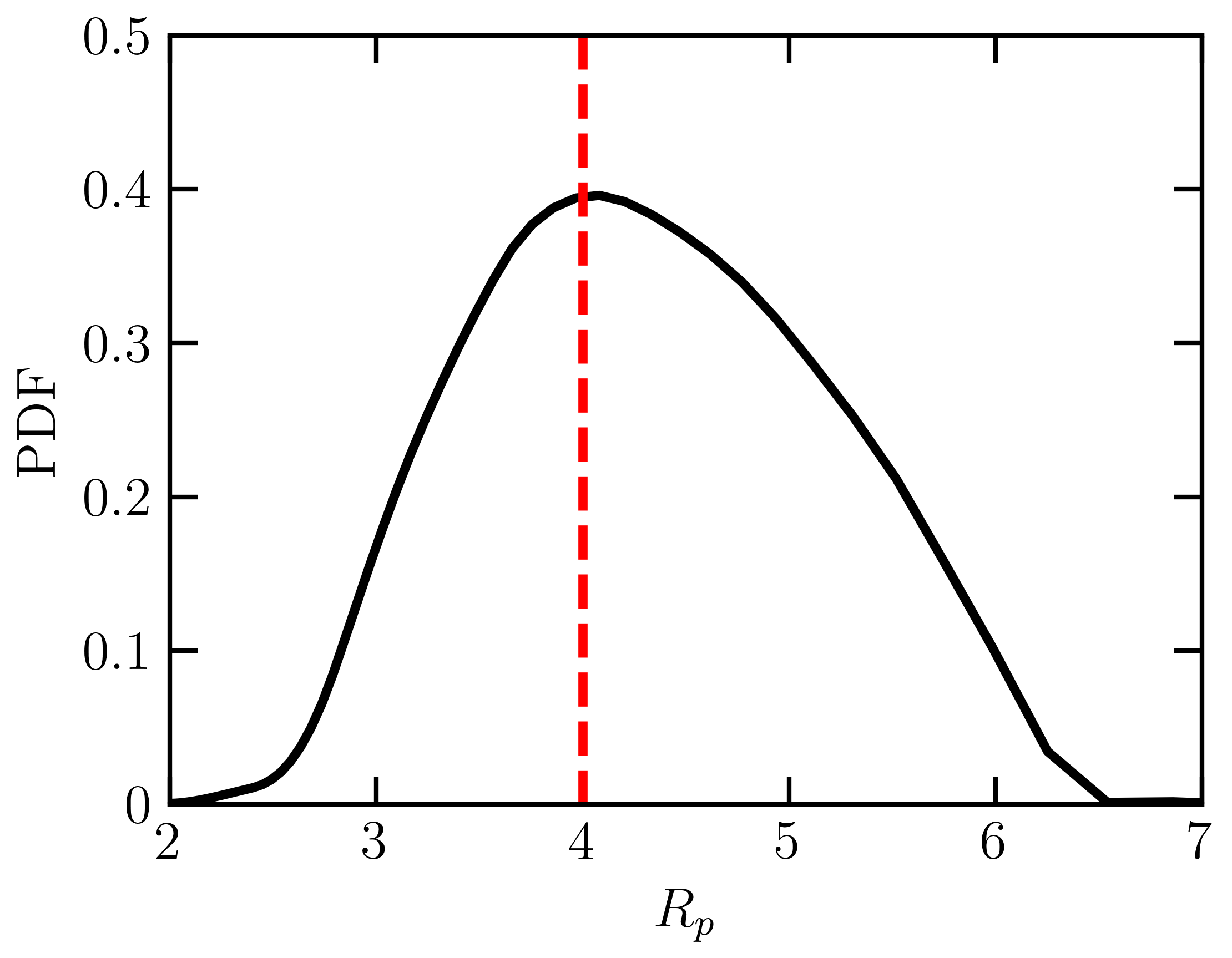}
    \caption{Probability density function (PDF) of the pore throat size ratio, $R_p = D/d_p$, in the randomly packed large particle bed (large particle volume fraction is 0.60).}
    \label{fig:pdf}
\end{figure}

Although the pore throat size ratio, $R_p$, varies substantially in randomly packed beds, it is possible to estimate a typical value for $R_p$ by considering the probability density function (PDF) of pore throat sizes, shown in Fig.~\ref{fig:pdf}, which is obtained following the Delaunay triangulation approach used previously ~\cite{gaoPercolationFineParticle2023a}. 
The PDF in Fig.~\ref{fig:pdf} has a broad peak  corresponding to $R_p = 4$, consistent with previous findings for randomly packed static beds~\cite{gaoPercolationFineParticle2023a}. Hence, we use $R_p = 4$ in Eq.~\ref{eq:Rscaling}.  

\begin{figure}
    \centering
        \includegraphics[width=\linewidth]{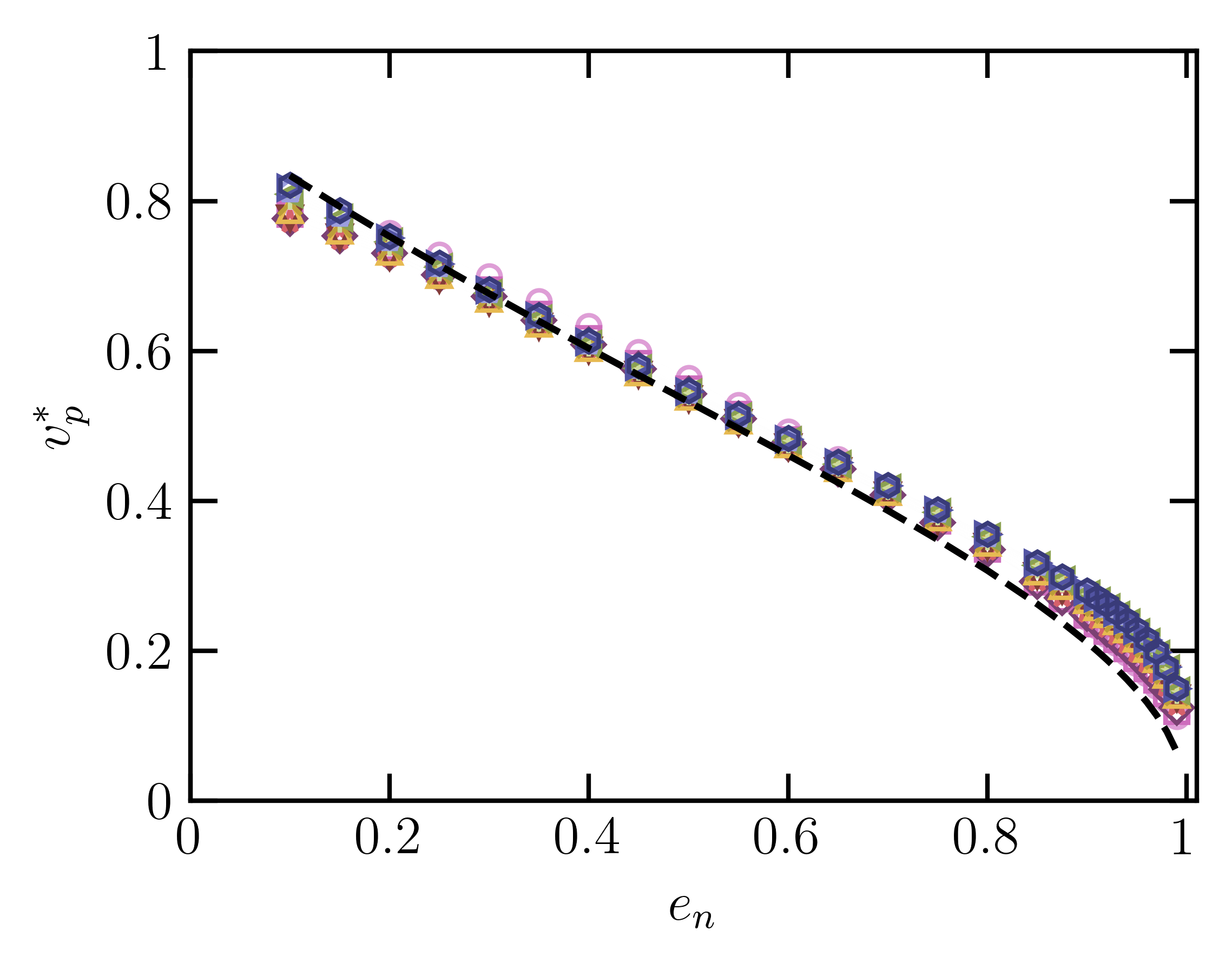}
    \caption{Size scaled percolation velocity $v_p^*$ vs $e_n$ for $7 \leq R \leq 50$  (symbols as in Fig.~\ref{fig:vp}). Dashed curve corresponds to $v_p/\sqrt{gD}$ from Eq.~\ref{eq:vpstarana} with $h'=1.7$.}
    \label{fig:vpanar}
\end{figure}

\begin{figure}
    \centering
        \includegraphics[width=\linewidth]{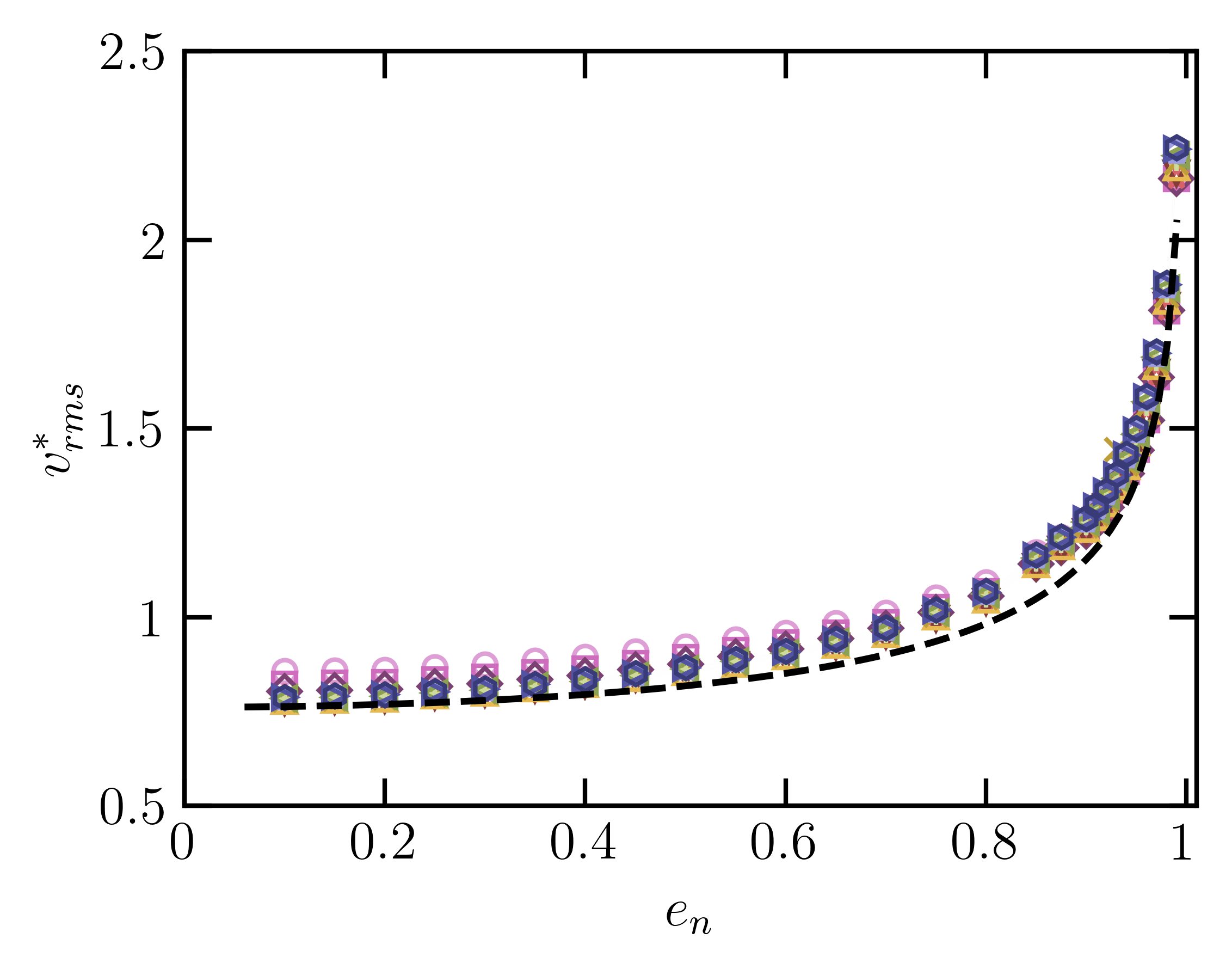}
    \caption{Size scaled fluctuation velocity   $v_{rms}^*$ vs $e_n$ for $7 \leq R \leq 50$  (symbols as in Fig.~\ref{fig:vp}). Dashed curve corresponds to $v_{rms}/\sqrt{gD}$ from Eq.~\ref{eq:vrmsanar2} with $\beta=0.39$.}
    \label{fig:vrmsanar2}
\end{figure}

To demonstrate the effectiveness of the scaling in Eq.~\ref{eq:Rscaling} in capturing the size ratio dependence, the DEM results in Fig.~\ref{fig:vp} and Fig.~\ref{fig:vrms} are replotted in  rescaled form in  Fig.~\ref{fig:vpanar} and Fig.~\ref{fig:vrmsanar2}, respectively.
The collapse across size ratios in Fig.~\ref{fig:vpanar} for the scaled percolation velocity has a coefficient of variation below 7\%. 
%
Similarly using the scaling of Eq.~\ref{eq:Rscaling} for  $v_{rms}$ collapses the data equally well in Fig.~\ref{fig:vrmsanar2}, with a coefficient of variation that remains below 8\%. Note that the $\sqrt{1-R_p/R}$ correction is relatively small, ranging from 0.65 at $R=7$ to 0.96 at $R=50$.


\subsection{Friction effects}
\label{sec:friction}

\begin{figure}
    \centering
        \includegraphics[width=\linewidth]{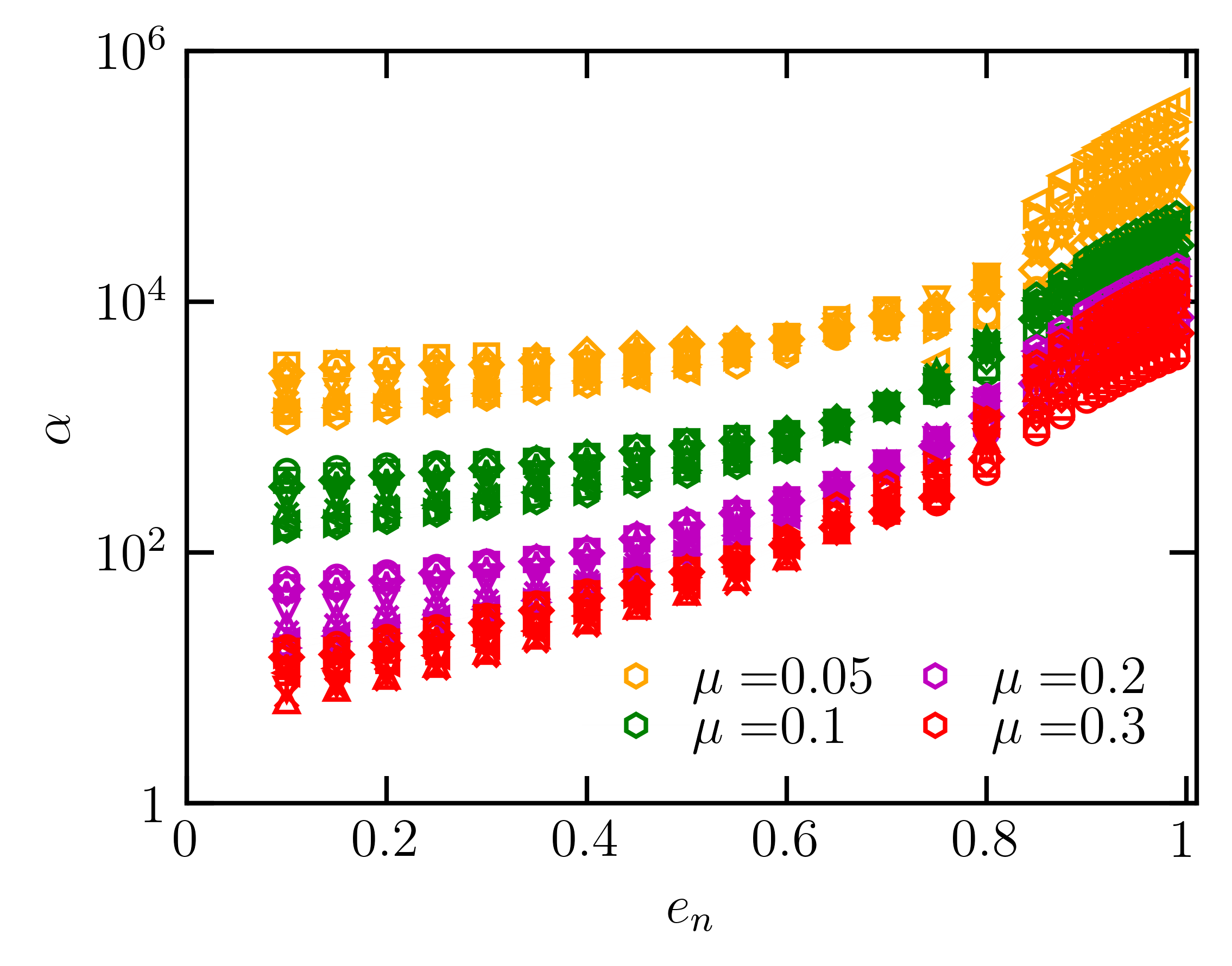}
    \caption{Trapping depth coefficient, $\alpha $, vs\ restitution coefficient, $e_n$, for different values of friction coefficient, $\mu$, (colors) and size ratios, $7 \leq R \leq 50$ (symbols as in Fig.~\ref{fig:vp}). }
    \label{fig:trapped}
\end{figure}

In addition to energy losses characterized by the restitution coefficient,  friction is the other key energy dissipation mechanism in granular systems. To examine their combined effect  on free-sifting particles, we vary the friction coefficient, $\mu$, from 0 to 0.3 and analyze its impact on $v_p$ and $v_{rms}$.  
First, however, it is important to note that the introduction of friction can cause fine particles to become trapped between two large particles~\cite{vyasImprovedVelocityVerletAlgorithm2024}. 
To quantify this behavior, we calculate the probability $P_p(\Delta z)$ that a fine particle percolates to a depth $\Delta z$ without becoming trapped for different values of $\mu$, $e_n$, and $R$. As in~\cite{gaoPercolationFineParticle2023a}, this can be expressed as
\begin{equation}
    P_p(\Delta z) = e^{-\Delta z/\alpha D},
    \label{eq:pplambda}
\end{equation}
where $\alpha D$ is the characteristic percolation depth, representing the distance a fine particle travels before being trapped in the large particle bed.
For frictionless and low $\mu$ systems ($\mu \leq 0.01$), $\alpha \to \infty$, indicating that fine particles percolate freely. As $\mu$ increases, $\alpha$ decreases, signifying more frequent trapping. As shown in Fig.~\ref{fig:trapped}, the trapping depth, quantified in terms of $\alpha$, decreases with increasing $\mu$ and decreasing $e_n$. 

In addition to friction, $e_n$ also influences trapping. Higher $e_n$ values lead to greater excitation and bouncing, thereby increasing $\alpha$ since particles only become trapped once they stop bouncing.
Although it is difficult to discern because of the overlap of the markers in Fig.~\ref{fig:trapped}, $\alpha$ increases with decreasing $R$, as larger fine particles are less likely to wedge between bed particles. 
For large friction values, with sufficient time, all particles  eventually become trapped, recalling that we consider fine particles that do not interact with one another and cannot dislodge previously trapped particles. 

\begin{figure}
    \centering
        \includegraphics[width=\linewidth]{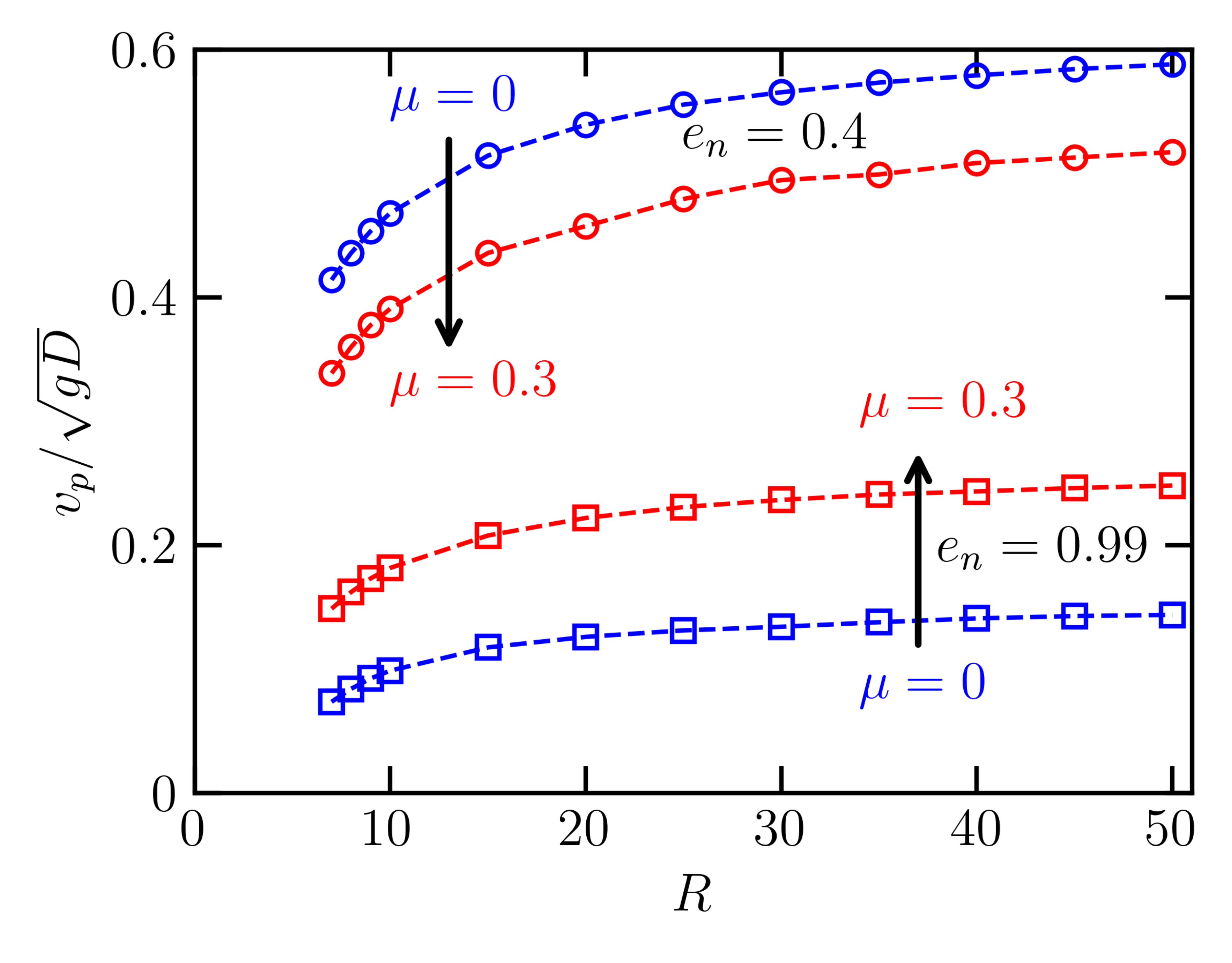}
    \caption{Scaled percolation velocity, $v_p/\sqrt{gD}$, vs size ratio, $R$, for friction coefficient, $\mu = 0$ (blue) and 0.3 (red), showing opposite responses for  restitution coefficients, $e_n = 0.4$ (circles) and $0.99$ (squares).}
    \label{fig:muvp}
\end{figure}

\begin{figure}
    \centering
        \includegraphics[width=\linewidth]{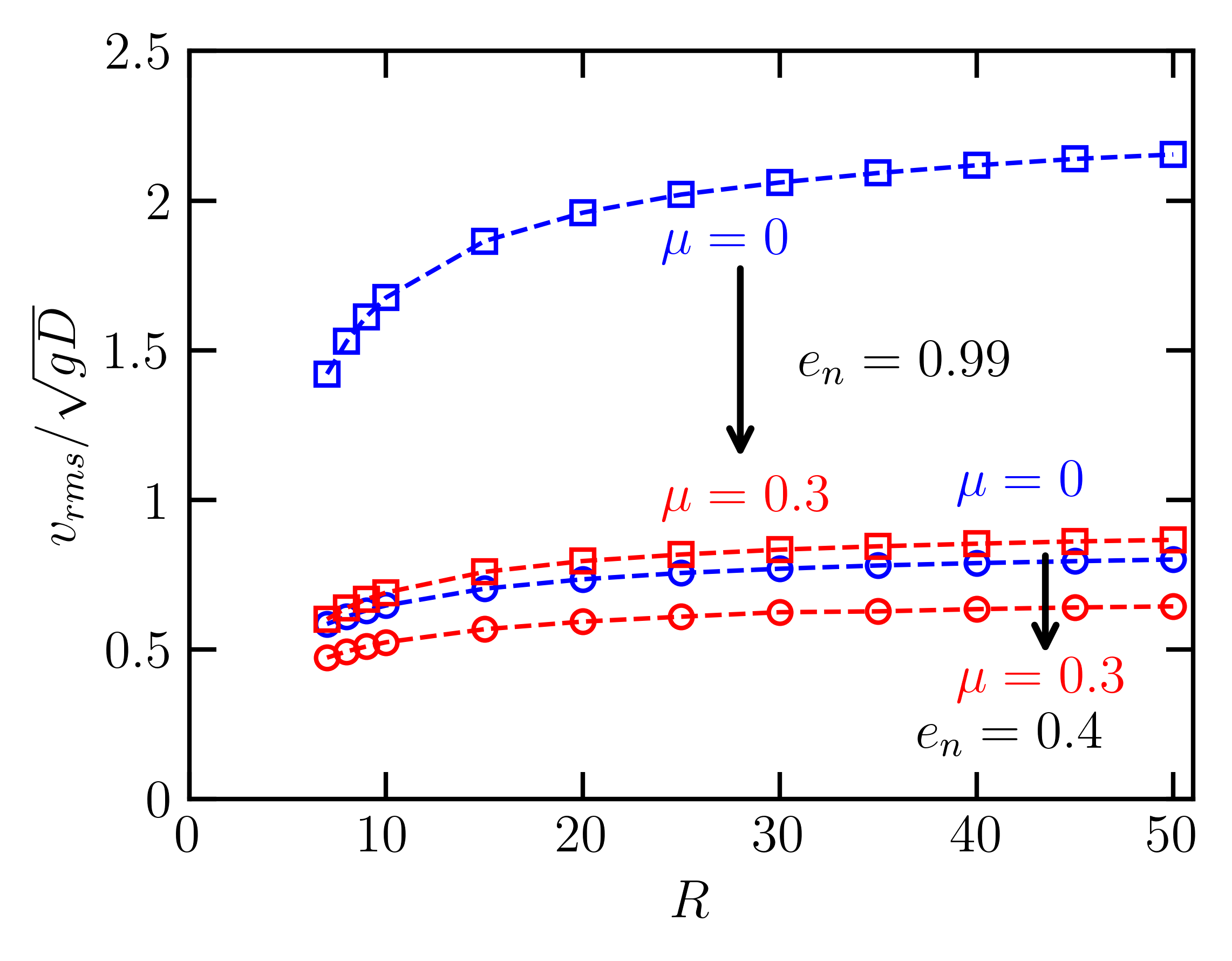}
    \caption{Scaled fluctuation velocity, $v_{rms}/\sqrt{gD}$, vs size ratio, $R$, decreases with increasing  friction coefficient (data shown for $\mu = 0$ (blue) and $0.3$ (red)) at low and high  restitution coefficients
    $e_n = 0.4$ (circles) and $0.99$ (squares).}
    \label{fig:muvrms}
\end{figure}

The combined effects of the restitution and friction dissipation mechanisms on $v_p$ and $v_{rms}$ are shown in Figs.~\ref{fig:muvp} and~\ref{fig:muvrms}, respectively, for $e_n = 0.4$ and $e_n = 0.99$ at $\mu = 0$ and $\mu = 0.3$. To prevent trapping from affecting $v_p$ and $v_{rms}$ measurements, we calculate steady-state values only for untrapped particles~\cite{gaoPercolationFineParticle2023a}. In Fig.~\ref{fig:muvp} for $e_n = 0.4$, $v_p$ decreases with increasing $\mu$, whereas for $e_n = 0.99$, $v_p$ increases. In contrast, in Fig.~\ref{fig:muvrms} $v_{rms}$ decreases with increasing $\mu$ for both cases, with a more pronounced reduction at $e_n = 0.99$.  
To explain, at high $e_n$, the absence of friction minimizes energy loss, leading to  greater bouncing and higher $v_{rms}$, which reduces $v_p$. Introducing friction enhances energy dissipation, substantially reducing $v_{rms}$ and enabling freer percolation, thereby increasing $v_p$. Conversely, at low $e_n$, $v_{rms}$ is already low, so introducing friction  only decreases it slightly. However, friction further inhibits sliding during collisions, thereby reducing $v_p$.  Data for other values of $e_n$ and $\mu$ fall between the curves shown in Figs.~\ref{fig:muvp} and \ref{fig:muvrms} but are omitted for clarity.

\begin{figure}[tbh]
    \centering
        \includegraphics[width=\linewidth]{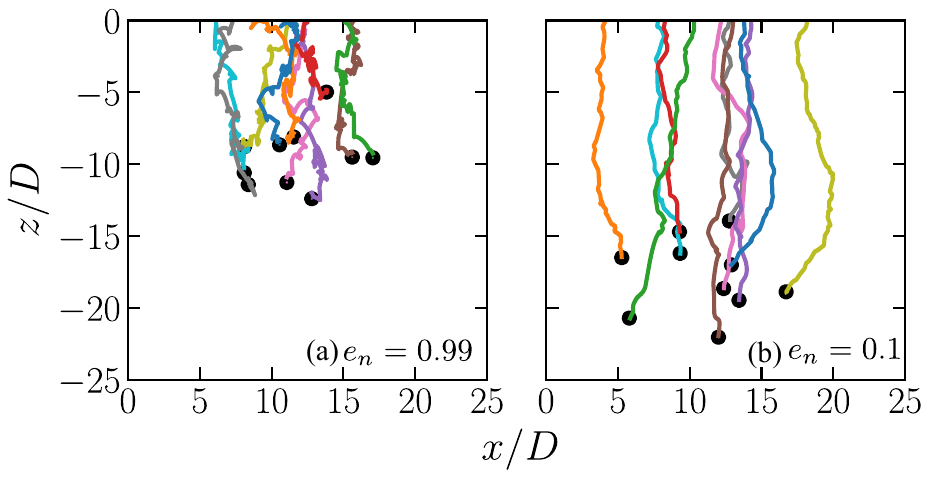}
    \caption{Sample unwrapped trajectories of 10 fine particles with (a) $e_n = 0.99$ (left) and (b) $e_n = 0.1$ (right) for $\mu = 0.3$ in a static bed of large particles with $R=50$ showing the difference in motion for low and high restitution coefficients.}
    \label{fig:traj2}
\end{figure}

\begin{figure}
    \centering
        \includegraphics[width=0.96\linewidth]{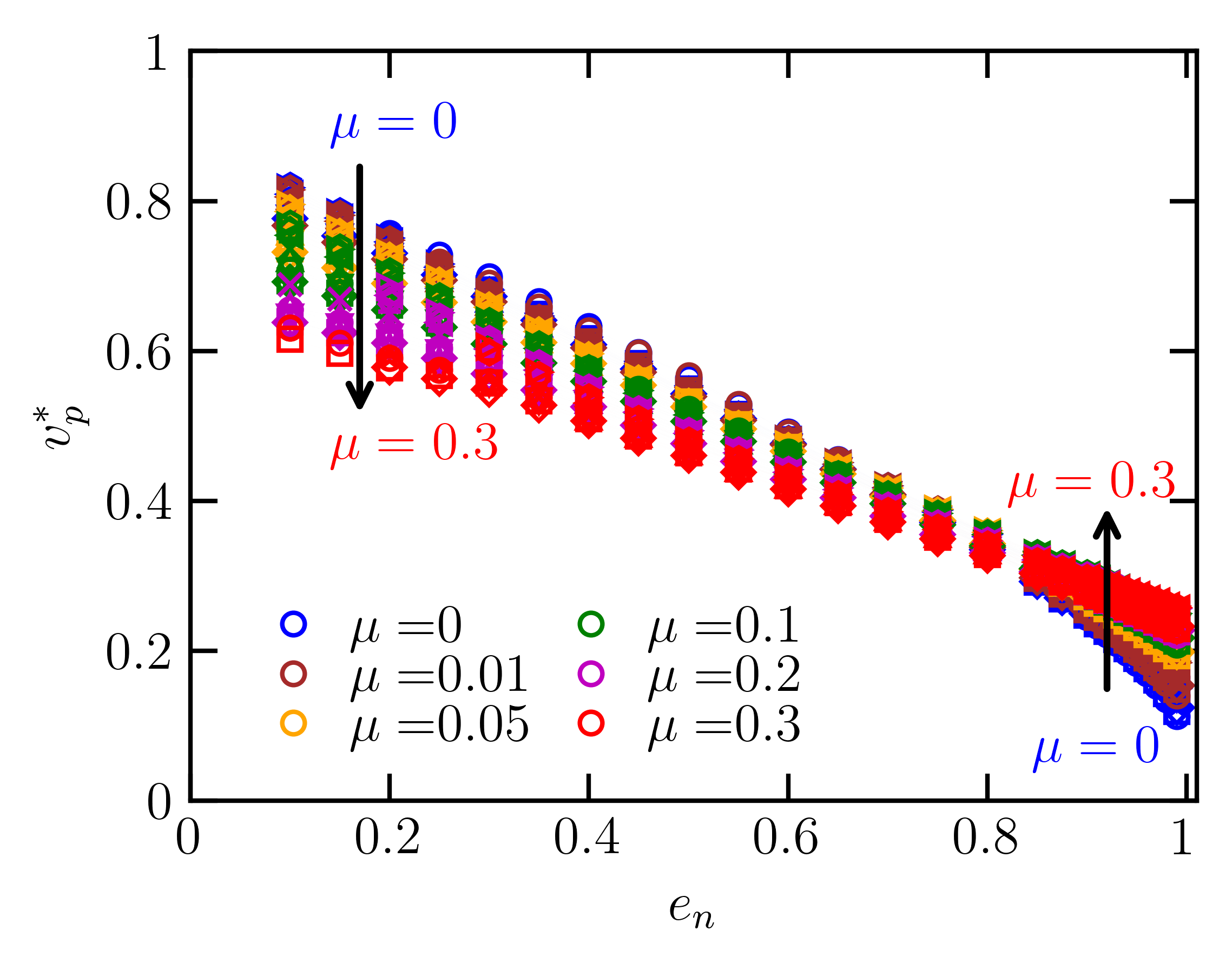}
    \caption{Scaled percolation velocity $v_p^*$ vs $e_n$ for different $\mu$ (colors) with $7 \leq R \leq 50$  as denoted by symbols in Fig.~\ref{fig:vp}.}
    \label{fig:muvpall}
\end{figure}

\begin{figure}
    \centering
        \includegraphics[width=0.96\linewidth]{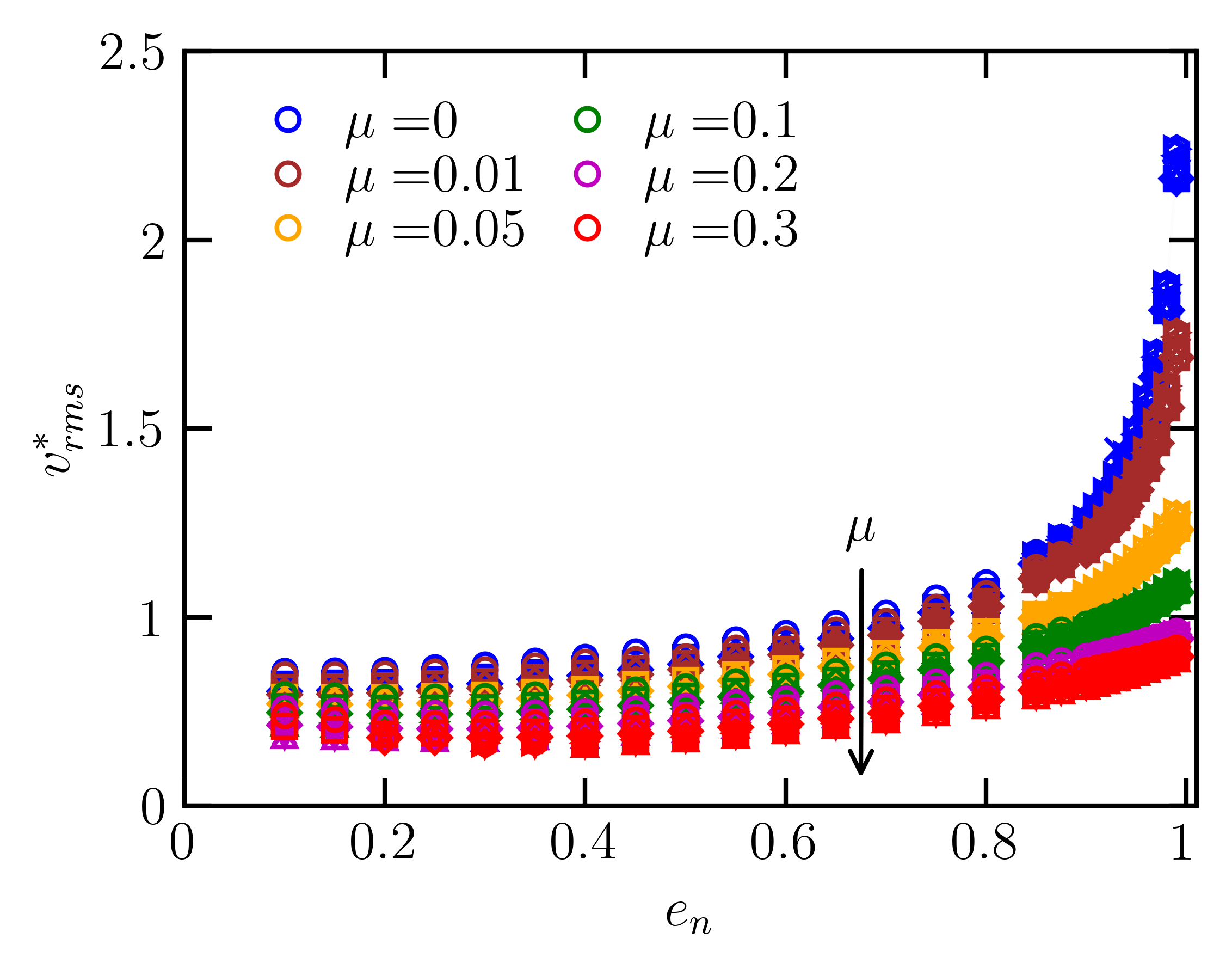}
    \caption{Scaled fluctuation velocity $v_{rms}^*$ vs $e_n$ for different $\mu$ (colors) with $7 \leq R \leq 50$  as denoted by symbols in Fig.~\ref{fig:vp}.}
    \label{fig:muvrmsall}
\end{figure}

The combined influence of restitution and friction is also evident in the fine-particle trajectories in Fig.~\ref{fig:traj2}  for $\mu = 0.3$ compared to the frictionless case in Fig.~\ref{fig:traj}. At high restitution ($e_n = 0.99$) the lack of friction results in minimal energy dissipation and strongly scattering trajectories. With friction, however, the additional frictional dissipation reduces the velocity fluctuations $v_{rms}$ and leads to more downwardly directed motion.  
At low restitution ($e_n = 0.1$), the trajectories remain governed by the gravity-dominated percolation process and are similar to the frictionless case. Yet, as fines percolate through the static bed, they slide against the bed particles, and friction increases the resistance to this sliding, thereby lowering the percolation velocity. This influence is clearly reflected in Figs.~\ref{fig:muvpall} and \ref{fig:muvrmsall}, which present the $R$-scaled percolation and fluctuation velocities across a range of $e_n$ and $\mu$. For $e_n \lesssim 0.8$, friction enhances sliding resistance, reducing both $v_p^*$ and $v_{rms}^*$. At higher restitution, however, the introduction of friction produces the opposite effect: $v_p^*$ increases because the stronger reduction in $v_{rms}^*$ suppresses random scattering, resulting in a more gravity-dominated percolation (Fig.~\ref{fig:traj2}~(a)). At $e_n \approx 0.8$, these opposing effects nearly balance, and the percolation velocity is largely independent of $\mu$.

An important observation in Figs.~\ref{fig:muvpall} and \ref{fig:muvrmsall} is the contrasting behavior for $R$-scaled fluctuation velocities depending on whether $v_{rms}^*$ is greater or less than one. For $e_n \gtrsim 0.8$ and $\mu \leq 0.05$, $v_{rms}^*$ rapidly rises above one (Fig.~\ref{fig:muvrmsall}) as $e_n$ increases, indicating strongly random motion, consistent with the trajectories in Fig.~\ref{fig:traj}(a) for $e_n = 0.99$. Under the same conditions, Fig.~\ref{fig:muvpall} shows a correspondingly steep reduction in percolation velocity.  
In contrast, for $e_n \gtrsim 0.8$ and $\mu \geq 0.1$, the additional energy dissipation from friction constrains the increase in $v_{rms}^*$, keeping its value below one (Fig.~\ref{fig:muvrmsall}). In this regime, the motion of fines is less random and more directed, as demonstrated by the trajectories in Fig.~\ref{fig:traj2} for $e_n = 0.99$. Consequently, the reduction in percolation velocity is more gradual (Fig.~\ref{fig:muvpall}), resembling the behavior seen at lower restitution ($e_n \lesssim 0.8$).  
These observations show that the transition from random to directed motion is governed by $v_{rms}^*$. 
When fluctuations are small, $v_{rms}^* \lesssim 1$,  motion is more directed and percolation velocity is higher. Conversely, when fluctuations are large, $v_{rms}^* \gtrsim 1$, motion becomes more random and the percolation velocity decreases.

In the next two sections, we analyze these two regimes separately and present simple conceptual models that capture their contrasting behaviors.

\subsection{Drude model for fluctuation-dominated percolation}
\label{sec:drude}

In Figs.~\ref{fig:vpanar} and \ref{fig:vrmsanar2} for the fluctuation-dominated regime, $v_{rms}^* \gtrsim 1$, a rapid reduction in percolation velocity is accompanied by a sharp increase in fluctuation velocity for $e_n \gtrsim 0.8$. A similar trend is seen in Figs.~\ref{fig:muvpall} and \ref{fig:muvrmsall} for low values of $\mu \leq 0.05$. Comparable behavior  in sheared beds has also been reported by Gao et al.~\cite{gaoVerticalVelocitySmall2024} at lower size ratios $7 \leq R \leq 10$. They observed that in the presence of shear the percolation velocity decreases not only with increasing $e_n$, but also with increasing shear rate, as higher shear rates generate larger $v_{rms}$. They further noted that $v_p \propto 1/v_{rms}$ and suggested that this behavior is analogous to that predicted by the Drude model for electron conduction in metals.  

To examine whether the Drude model also applies to static beds, we compare our results with those for a sheared bed~\cite{gaoVerticalVelocitySmall2024}, where  the vertical component of the RMS velocity, $v_{z,\mathrm{rms}}$, is used to characterize fluctuations. To compare with our results here we assume that at high shear rates all three components of $v_{rms}$ are equal so that, $v_{rms} = \sqrt{3}\, v_{z,rms}$.

Since in static beds the value of $v_{rms}^*$ is constrained by $e_n$, we also consider a third configuration in which fines percolate through a bed of randomly excited large particles. This system retains the same geometry, material properties, size ratios, and packing fraction as the static bed, but introduces random excitation of the large particles using a Langevin thermostat implemented in LAMMPS~\cite{thompsonLAMMPSFlexibleSimulation2022, Schneider1978MoleculardynamicsSO} with a damping parameter of 1~s. For this system, we study two restitution coefficients, $e_n = 0.5$ and $0.9$, with $7 \leq R \leq 50$ and $\mu = 0$. The excitation velocity of the large particles is varied from $0.5\,\mathrm{m/s}$ to $9\,\mathrm{m/s}$ to obtain  fine-particle fluctuation velocities between $0.2\,\mathrm{m/s}$ and $1.4\,\mathrm{m/s}$, comparable to those reported in the sheared bed~\cite{gaoVerticalVelocitySmall2024}. Similar to the static configuration, the velocities of percolating fines reach steady state within one second, after which both $v_p$ and $v_{rms}$ are obtained as time- and ensemble-averaged quantities.

\begin{figure}[tbh]
    \begin{subfigure}{\linewidth}
        \includegraphics[width=\linewidth]{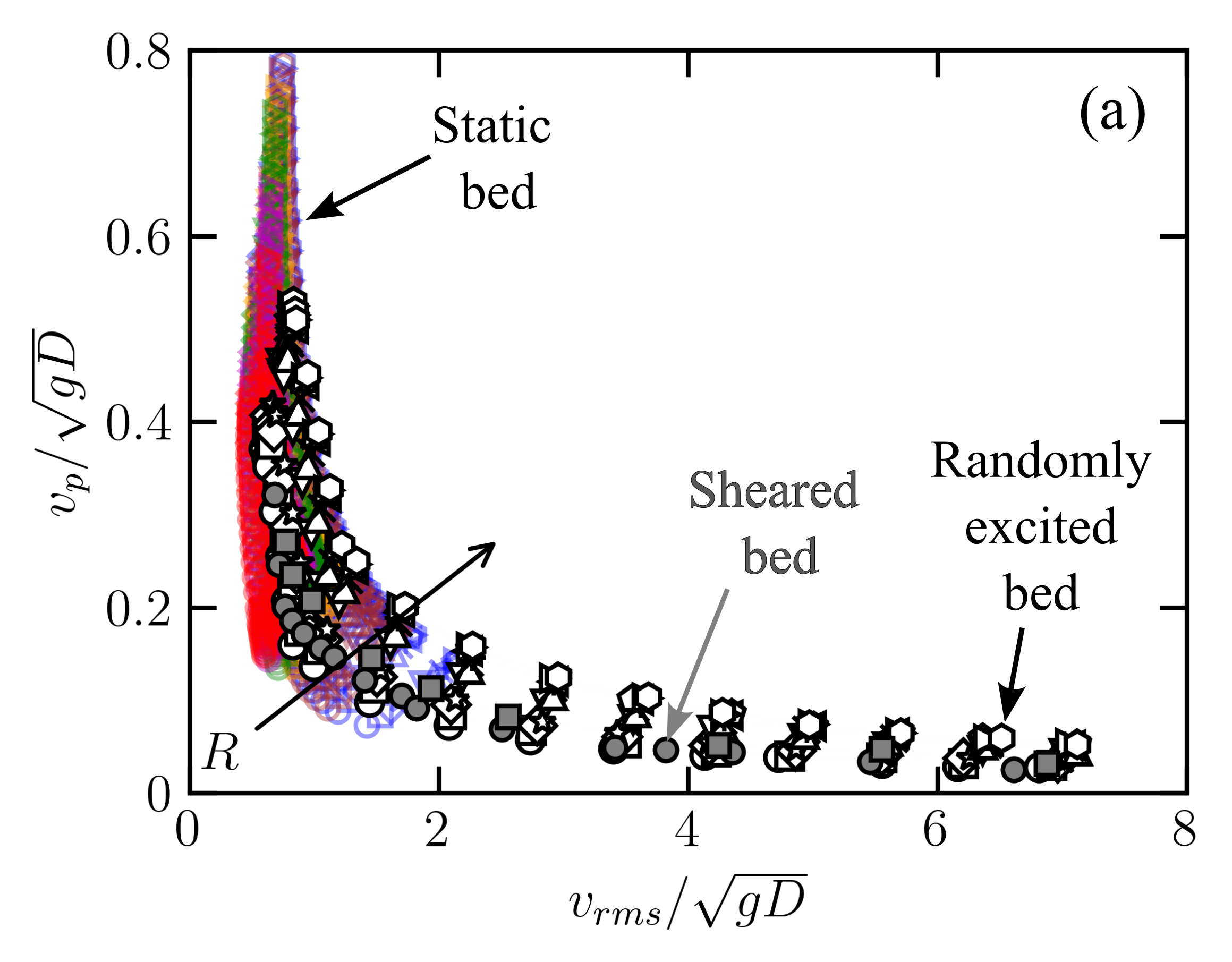}
        \label{fig:vpvrmsall_a}
    \end{subfigure}
    \begin{subfigure}{\linewidth}
        \includegraphics[width=\linewidth]{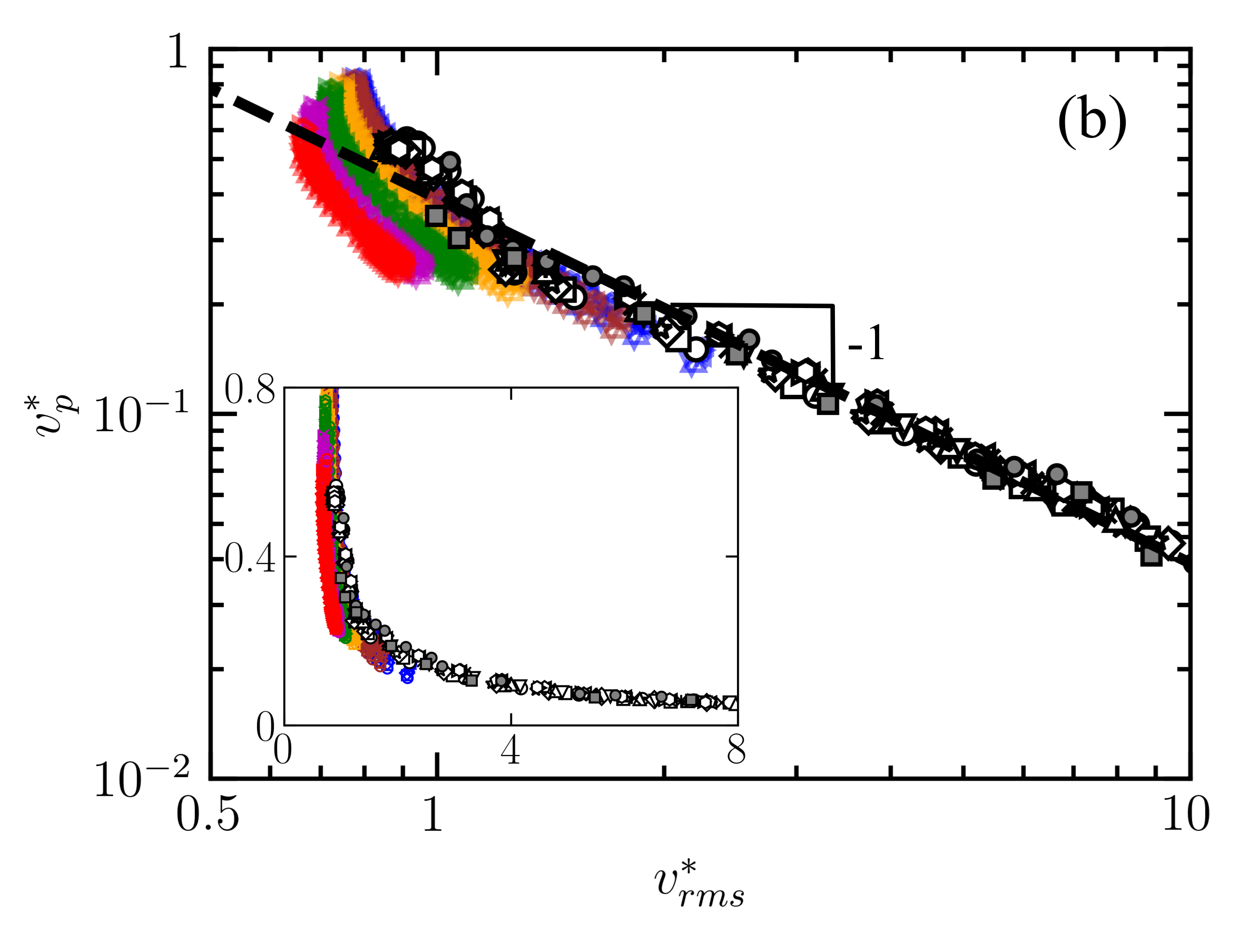}
        \label{fig:vpvrmsall_b}
    \end{subfigure}
    \caption{ (a) Inverse relationship between $v_p$ and $v_{rms}$ for static (colored symbols, varying $\mu$ as in Fig.~\ref{fig:muvpall}), sheared (grey symbols), and randomly excited (open black symbols) beds, with $7 \leq R \leq 50$ and $0.1 \leq e_n \leq 0.99$. Symbols represent $R$ values as in Fig.~\ref{fig:vp}. (b)  Same data scaled by $R$, shown on a logarithmic scale, with inset on a linear scale.}
    \label{fig:vpvrmsall}
\end{figure}

The inverse relationship between $v_p$ and $v_{rms}$ for static, sheared, and randomly excited beds is shown in Fig.~\ref{fig:vpvrmsall}. Figure~\ref{fig:vpvrmsall}(a) demonstrates that the inverse correlation between percolation velocity and fluctuation velocity holds across all excitation mechanisms at large $v_{rms}$, indicating that the relationship is largely independent of how fine particle velocity fluctuations are generated. Figure~\ref{fig:vpvrmsall}(b) shows that $R$-scaling collapses the data for $v_{rms}^* \gtrsim 1$,  independent of the fine particle excitation mechanisms. The log--log representation in Fig.~\ref{fig:vpvrmsall}(b) emphasizes the robustness of this trend at high $v_{rms}^*$ but also shows clear contrasting behavior at low fluctuation velocities, where the inverse correlation does not hold. The inset shows the collapse of the data on a linear scale.

The Drude model for electron conduction in metals mirrors this behavior, as described previously for sheared beds~\cite{gaoVerticalVelocitySmall2024}. The inverse relationship, \( v_p \propto 1/v_{rms} \), for fine-particle mobility parallels how increased lattice scattering reduces electron drift velocity in the Drude framework under an applied electric field \( E \).
In the Drude model, the mean momentum of the electron is given by $p = m v = qE\tau$,
where \( m \) and \( q \) are the electron mass and charge, respectively, and \( \tau \) is the mean time between collisions with atoms in the lattice. Drawing an analogy to fine particle percolation in the limit of a point particle ($R \to \infty$), we map the terms in the Drude model as follows: the electron drift velocity $v$ is replaced by the percolation velocity $v_p$, the driving acceleration $qE/m$ is replaced by gravitational acceleration $g$, and the mean free time $\tau$ is represented as $\beta D / v_{rms}$, where $\beta D$ represents the mean free path  of a point particle, which must be of order $D$. 
This analogy then gives the relation
\begin{equation}
v_p = \frac{\beta gD}{v_{rms}},
\label{eq:drude}
\end{equation}
which captures the inverse dependence of the percolation (drift) velocity on the fine particle's fluctuation (excitation) velocity for a point particle.
Incorporating $R$-scaling by substituting $\beta D(1-R_p/R)$ for $\beta D$ to account for finite fine particle size effects on the mean free path as discussed in regard to Eq.~\ref{eq:clearance} and Eq.~\ref{eq:Rscaling}, gives
\begin{equation}
    v_p = \frac{\beta gD}{v_{rms}}\left(1 - \frac{R_p}{R}\right)  \text{ or } v_p^* = \frac{\beta}{v_{rms}^*}.
\label{eq:drude2}
\end{equation}

In Fig.~\ref{fig:vpvrmsall}(b), we compare the DEM results with predictions from the Drude model. When $v_{rms}^* \gtrsim 1$, the Drude model agrees well with the DEM data using a fitted value of $\beta = 0.39$. Since $\beta D$ represents the mean free path  of a point particle, this corresponds to the characteristic pore size encountered by a fine particle traversing a static bed, which lies between the typical void diameters of tetrahedral ($0.225D$) and octahedral ($0.414D$) bed particle configurations, indicating that the model reflects realistic geometrical constraints of granular packings. 

In the Drude model, the mean free time is governed by the temperature, which determines the frequency of electron-lattice collisions. In contrast, for granular systems, the mean free time, represented as the ratio of mean free path to fluctuation velocity, $\beta D/v_{rms}$, is governed by the particle dynamics and the geometry of the bed. By combining this granular analog of the Drude model with a power balance argument, we estimate the dependence of $v_{rms}$ on $e_n$, capturing how energy dissipation due to inelastic collisions influences fine particle excitation. In a frictionless system, where energy losses stem solely from inelastic collisions, a simple power balance can be expressed as 
\begin{equation}
    mgv_p = \frac{1}{2} m v_{rms}^2 (1-e_n^2) n,
\end{equation}
where $n$ is the collision frequency. Solving for $v_{rms}$, the percolation velocity can be approximated as:
\begin{equation}
    v_p = n v_{rms}^2 \frac{(1-e_n^2)}{2g}.
    \label{eq:powerbalance}
\end{equation}
The collision frequency for high $v_{rms}$ systems can be estimated as $n = v_{rms}/\beta D$, giving
\begin{equation}
    v_p = \frac{v_{rms}^3 (1-e_n^2)}{2 \beta gD}.
    \label{eq:vpvrmsdrude}
\end{equation}
Combining this with Eq.~\ref{eq:drude} gives an expression for the fluctuation velocity as:
\begin{equation}
    \frac{v_{rms}}{\sqrt{gD}} = \sqrt{\beta} \left(\frac{2}{1-e_n^2}\right)^{1/4}.
    \label{eq:vrmsdrude}
\end{equation}  
To account for the size ratio, we apply the $\sqrt{1-R_p/R}$ correction to the left hand side so that
\begin{equation}
    v_{rms}^* = \sqrt{\beta} \left(\frac{2}{1-e_n^2}\right)^{1/4}.
    \label{eq:vrmsanar2}
\end{equation}
As shown in Fig.~\ref{fig:vrmsanar2}, this scaling captures the dependence of $v_{rms}^*$ on $e_n$ in a static-bed.
While the agreement in Fig~\ref{fig:vrmsanar2} is expected for $v_{rms}^* \gtrsim 1$, it is surprising that the model also fits the data reasonably well for $v_{rms}^* \lesssim 1$. 

Returning to Fig.~\ref{fig:vpvrmsall}(b), the spread of the data for $v_p^*$ broadens for $v_{rms}^* \lesssim 1$. This is consistent with the physical picture: under these conditions, fine-particle motion is gravity-dominated and violates the Drude model assumption of random motion, as evident from the trajectory plots in Figs.~\ref{fig:traj} and \ref{fig:traj2}, which show a more vertically biased descent. Thus, for typical values of $e_n$ and $\mu$  in physical systems,  Drude-like behavior will not occur in static bed percolation.
This motivates the development of a new model, introduced in the following section, that captures percolation dynamics in the low-$v_{rms}$ regime by explicitly accounting for gravity-driven motion and energy dissipation during successive collisions of a fine particle with bed particles.

\subsection{Staircase model for gravity-dominated percolation}
\label{sec:staircase}

\begin{figure}
    \centering
        \includegraphics[width=0.6\linewidth]{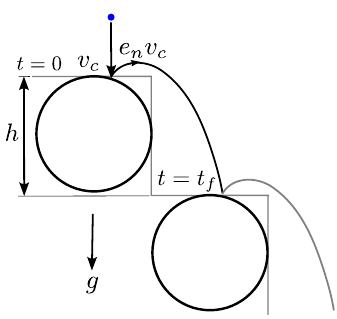}
    \caption{Schematic showing a simple model of a fine particle bouncing down a staircase of bed particles.}
    \label{fig:staircase}
\end{figure}

Knowing that low restitution minimizes excitation of a fine particle as it percolates through the bed, we propose a simplified representation of a single fine particle percolating through a static bed in the gravity dominated regime where $v_{rms}^* \lesssim 1$. This description, referred to as the ``staircase model,'' is shown schematically in Fig.~\ref{fig:staircase}.

At steady state, a fine particle with mass $m$ and typical vertical collision velocity $v_z = v_c$ bounces off the surface of a large particle with a reduced rebound velocity of $e_n v_c$. The fine particle then follows a parabolic trajectory before colliding again at a vertical distance $h \sim D$ below the previous collision location. The steady-state collision velocity $v_c$ can be estimated by balancing the gravitational potential energy gained between impacts, $mgh$, with the energy dissipated by inelastic loss, characterized by the restitution coefficient $e_n$, during each collision.

\begin{equation}
    \frac{1}{2} m v_c^2 (1-e_n^2)  = m g h,
    \label{eq:energybalance}
\end{equation}
which gives,
\begin{equation}
    v_c = \sqrt{\frac{2gh}{1-e_n^2}}.
    \label{eq:vc}
\end{equation}

Since the fine particle follows a parabolic trajectory with time of flight $t_f$ between consecutive impacts (steps), the associated average percolation velocity through the bed is $v_p = h/t_f$. By solving the quadratic kinematic equation for the time of flight for the parabolic trajectory after each collision, the percolation velocity, $v_p$, can be expressed as:
\begin{equation}
    v_p = \sqrt{\frac{gh}{2}}\sqrt{\frac{1-e_n}{1+e_n}}.
    \label{eq:vpana}
\end{equation}
Note that the staircase model ignores the additional dissipative effects associated with friction.
When $e_n$ approaches zero, the percolation velocity approaches $\sqrt{gh/2}$, which corresponds to the particle falling from step to step without bouncing. Conversely as $e_n$ approaches one, the percolation velocity approaches zero, as most of the kinetic energy is retained and the particle bounces higher and with a longer $t_f$, resulting in minimal downward percolation through the bed. 
In contrast to the Drude model, where the fine particle undergoes multiple collisions as it bounces in a bed of large particles and $v_p$ and $\tau$ are strongly influenced by $v_{rms}$, the staircase model assumes a single collision between steps and is governed solely by the parabolic motion under gravity between successive impacts, which depends on $e_n$. 

The step height $h$ represents the vertical distance between collisions and should be proportional to the bed particle diameter as well as the mean free path.  Consequently, we express $h$ as 
\begin{equation}
    h= h' D \left(1-\frac{R_p}{R}\right),
\end{equation}
where $h'D$ represents the characteristic distance a point particle ($R \to \infty$) falls between collisions.  Substituting this expression into Eq.~\ref{eq:vpana} and rearranging gives 
\begin{equation}
    v_p^* = \sqrt{\frac{h'}{2D}}\sqrt{\frac{1-e_n}{1+e_n}}.
    \label{eq:vpstarana}
\end{equation}
Equation~\ref{eq:vpstarana} matches the data in Fig.~\ref{fig:vpanar} well using $h' = 1.7D$. The agreement is best at higher $R$ for $e_n \lesssim 0.8$ ($v_{rms}^* \lesssim 1$), whereas for $e_n \gtrsim 0.8$ the predictions align more closely with the DEM results at lower $R$, where $v_{rms}^*$ is also relatively small as seen in Fig.~\ref{fig:vrms}.
Hence, Eq.~\ref{eq:vpstarana} can be used to estimate the percolation velocity for a given size ratio and restitution coefficient.
While the staircase model could be further refined by incorporating effects like large particle curvature, energy dissipation due to friction during interparticle collisions, and rolling and sliding motions of fine particles, it is striking how well this simple model captures the observed percolation behavior in the regime where $v_{rms}^* \lesssim 1$.

\section{Conclusions}

In this study, we investigate the percolation of non-cohesive spherical fine particles in randomly packed granular beds in the absence of interstitial fluids. We examine the influence of different interaction parameters, the coefficient of restitution ($e_n$), the friction coefficient ($\mu$), and the large-to-fine diameter ratio ($R$), on fine particle percolation and fluctuation velocities.
In static beds, $e_n$ strongly influences percolation: as $e_n$ increases, percolation velocity, $v_p$, decreases while fluctuation velocity, $v_{rms}$, increases. Two regimes emerge based on the fluctuation velocity, $v_{rms}$. When $v_{rms}/\sqrt{gD} \lesssim 1$ percolation is gravity-dominated and step-like, whereas when $v_{rms}/\sqrt{gD} \gtrsim 1$, random  collisional dynamics dominate, leading to an inverse $v_p$-$v_{rms}$ relationship. 
This inverse scaling applies not only to static beds but equally well to sheared, and randomly excited beds. 
However, even in these externally excited system when the excitation is weak, the percolation will fall into the gravity dominated regime.
While the Drude model captures the high-fluctuation regime, it fails at capturing  behavior for $v_{rms} \lesssim 1$, where a staircase model better describes the gravitational descent of fines.

The size ratio $R$ also influences $v_p$ and $v_{rms}$, although not as strongly as $e_n$. As $R$ increases, both $v_p$ and $v_{rms}$ initially rise but then plateau. This behavior is attributed to smaller particles navigating pore throats more readily, enhancing percolation and thereby reducing their collision frequency with the bed particles, which in turn decreases energy dissipation and results in increased fluctuation velocities.
Friction affects $v_p$ and $v_{rms}$ differently depending on $e_n$. For restitution coefficients $e_n \lesssim 0.8$, increasing friction reduces $v_p$ due to inhibited sliding  of fine particles over larger ones  during collisions. At higher restitution coefficients ($e_n \gtrsim 0.8$), friction decreases $v_{rms}$, which, in turn, increases $v_p$ as particles lose energy and percolate more quickly.

While the present study focuses exclusively on dry systems without interstitial fluids, simple force-balance estimates allow us to identify particle sizes above which fluid and cohesive effects are negligible.
In air, a fine glass or sand particle larger than roughly 200\,\textmu m can be considered to percolate freely under gravity, as fluid drag is negligible. In water, this threshold increases to approximately 1\,mm, due to buoyancy and viscous resistance reducing the effective gravitational acceleration.
Cohesive forces, particularly van der Waals interactions, are another important consideration for very fine particles. For dry glass beads, such effects become significant below about 150\,\textmu m, meaning that for particles larger than 200\,\textmu m both cohesion and fluid effects can be safely ignored in air, but for particles smaller than 200\,\textmu m these effects must be considered.

Although this study has untangled the intertwined effects of restitution, friction, and size ratio over a wide parameter space, $0.1\leq e_n \leq 0.99$, $0 \leq \mu \leq 0.3$ and $7\leq R \leq 50$, many question remain. First and foremost is to explore percolation of cohesive and non-cohesive interacting fine particles (rather than the individual fine particles considered here), which would reflect more typical segregation conditions.  Another critical gap in understanding, particularly for industrial and geophysical applications, is that of interacting fine particles in sheared and randomly excited granular beds for both low concentrations of fine particles, where fine particles mostly interact with large particles collisionally, and for high enough concentrations of fine particles that the fine particles act as a ``continuous phase'' with a ``dispersed phase'' of large particles suspended within them.  

\section{Funding Information}

This material is based upon work supported by the National Science Foundation under Grant No. CBET-2203703.

\bibliographystyle{elsarticle-num} 
\bibliography{exported}

\begin{thebibliography}{10}
\expandafter\ifx\csname url\endcsname\relax
  \def\url#1{\texttt{#1}}\fi
\expandafter\ifx\csname urlprefix\endcsname\relax\def\urlprefix{URL }\fi
\expandafter\ifx\csname href\endcsname\relax
  \def\href#1#2{#2} \def\path#1{#1}\fi

\bibitem{duranIntroduction2000}
J.~Duran, Introduction, in: Sands, {{Powders}}, and {{Grains}}: {{An
  Introduction}} to the {{Physics}} of {{Granular Materials}}, Springer, New
  York, NY, 2000, pp. 1--18.
\newblock \href {https://doi.org/10.1007/978-1-4612-0499-2}
  {\path{doi:10.1007/978-1-4612-0499-2}}.

\bibitem{sarkarRoleForcesGoverning2017}
S.~Sarkar, R.~Mukherjee, B.~Chaudhuri, On the role of forces governing
  particulate interactions in pharmaceutical systems: {{A}} review,
  International Journal of Pharmaceutics 526~(1) (2017) 516--537.
\newblock \href {https://doi.org/10.1016/j.ijpharm.2017.05.003}
  {\path{doi:10.1016/j.ijpharm.2017.05.003}}.

\bibitem{johnParticleBreakageConstruction2023}
N.~J. John, I.~Khan, S.~Kandalai, A.~Patel, Particle breakage in construction
  materials: {{A}} geotechnical perspective, Construction and Building
  Materials 381 (2023) 131308.
\newblock \href {https://doi.org/10.1016/j.conbuildmat.2023.131308}
  {\path{doi:10.1016/j.conbuildmat.2023.131308}}.

\bibitem{horabikParametersContactModels2016}
J.~Horabik, M.~Molenda, Parameters and contact models for {{DEM}} simulations
  of agricultural granular materials: {{A}} review, Biosystems Engineering 147
  (2016) 206--225.
\newblock \href {https://doi.org/10.1016/j.biosystemseng.2016.02.017}
  {\path{doi:10.1016/j.biosystemseng.2016.02.017}}.

\bibitem{floreAspectsGranulationChemical2009}
K.~Flore, M.~Schoenherr, H.~Feise, Aspects of granulation in the chemical
  industry, Powder Technology 189~(2) (2009) 327--331.
\newblock \href {https://doi.org/10.1016/j.powtec.2008.04.010}
  {\path{doi:10.1016/j.powtec.2008.04.010}}.

\bibitem{meloshMechanicsLargeRock1987}
H.~J. Melosh, The mechanics of large rock avalanches, in: J.~E. Costa, G.~F.
  Wieczorek (Eds.), Debris {{Flows}}/{{Avalanches}}, Vol.~7, Geological Society
  of America, 1987, p.~0.
\newblock \href {https://doi.org/10.1130/REG7-p41}
  {\path{doi:10.1130/REG7-p41}}.

\bibitem{grayRapidGranularAvalanches2003}
J.~M. N.~T. Gray, Rapid granular avalanches, in: K.~Hutter, N.~Kirchner (Eds.),
  Dynamic response of granular and porous materials under large and
  catastrophic deformations, Springer, Berlin, Heidelberg, 2003, pp. 3--42.
\newblock \href {https://doi.org/10.1007/978-3-540-36565-5_1}
  {\path{doi:10.1007/978-3-540-36565-5_1}}.

\bibitem{sanchezSimulatingAsteroidRubble2011}
P.~S{\'a}nchez, D.~J. Scheeres, Simulating asteroid rubble piles with a
  self-gravitating soft-sphere distinct element method model, The Astrophysical
  Journal 727~(2) (2011) 120.
\newblock \href {https://doi.org/10.1088/0004-637X/727/2/120}
  {\path{doi:10.1088/0004-637X/727/2/120}}.

\bibitem{Fang2024CloggingTO}
W.~Fang, S.~Chen, S.~Li, I.~Zuriguel, Clogging transition of granular flow in
  porous structures, Physical Review Research (2024).
\newblock \href {https://doi.org/10.1103/physrevresearch.6.033046}
  {\path{doi:10.1103/physrevresearch.6.033046}}.

\bibitem{Sharma1987FinesMI}
M.~Sharma, Y.~Yortsos, Fines migration in porous media, AIChE Journal 33 (1987)
  1654--1662.
\newblock \href {https://doi.org/10.1002/aic.690331009}
  {\path{doi:10.1002/aic.690331009}}.

\bibitem{Zhu2009}
H.~P. Zhu, M.~Rahman, A.~B. Yu, J.~Bridgwater, P.~Zulli, Effect of particle
  properties on particle percolation behaviour in a packed bed, Minerals
  Engineering 22~(11) (2009) 961--969.
\newblock \href {https://doi.org/10.1016/j.mineng.2009.03.002}
  {\path{doi:10.1016/j.mineng.2009.03.002}}.

\bibitem{shinEffectBallSize2013}
H.~Shin, S.~Lee, H.~Suk~Jung, J.-B. Kim, Effect of ball size and powder loading
  on the milling efficiency of a laboratory-scale wet ball mill, Ceramics
  International 39~(8) (2013) 8963--8968.
\newblock \href {https://doi.org/10.1016/j.ceramint.2013.04.093}
  {\path{doi:10.1016/j.ceramint.2013.04.093}}.

\bibitem{hattoriRockFragmentationParticle1999}
I.~Hattori, H.~Yamamoto, Rock fragmentation and particle size in crushed zones
  by faulting, The Journal of Geology 107~(2) (1999) 209--222.
\newblock \href {https://doi.org/10.1086/314343} {\path{doi:10.1086/314343}}.

\bibitem{crostaFragmentationValPola2007}
G.~B. Crosta, P.~Frattini, N.~Fusi, Fragmentation in the {{Val Pola}} rock
  avalanche, {{Italian Alps}}, Journal of Geophysical Research: Earth Surface
  112~(F1) (2007).
\newblock \href {https://doi.org/10.1029/2005JF000455}
  {\path{doi:10.1029/2005JF000455}}.

\bibitem{shekunovParticleSizeAnalysis2007}
B.~Y. Shekunov, P.~Chattopadhyay, H.~H.~Y. Tong, A.~H.~L. Chow, Particle {{Size
  Analysis}} in {{Pharmaceutics}}: {{Principles}}, {{Methods}} and
  {{Applications}}, Pharmaceutical Research 24~(2) (2007) 203--227.
\newblock \href {https://doi.org/10.1007/s11095-006-9146-7}
  {\path{doi:10.1007/s11095-006-9146-7}}.

\bibitem{tongNumericalStudyEffects2010}
Z.~B. Tong, R.~Y. Yang, K.~W. Chu, A.~B. Yu, S.~Adi, H.~K. Chan, Numerical
  study of the effects of particle size and polydispersity on the agglomerate
  dispersion in a cyclonic flow, Chemical Engineering Journal 164~(2) (2010)
  432--441.
\newblock \href {https://doi.org/10.1016/j.cej.2009.11.027}
  {\path{doi:10.1016/j.cej.2009.11.027}}.

\bibitem{Umbanhowar2019ModelingSI}
P.~Umbanhowar, R.~M. Lueptow, J.~Ottino, Modeling segregation in granular
  flows., Annual Review of Chemical and Biomolecular Engineering 10 (2019)
  129--153.
\newblock \href {https://doi.org/10.1146/annurev-chembioeng-060718-030122}
  {\path{doi:10.1146/annurev-chembioeng-060718-030122}}.

\bibitem{fan_modelling_2014}
Y.~Fan, C.~P. Schlick, P.~B. Umbanhowar, J.~M. Ottino, R.~M. Lueptow, Modelling
  size segregation of granular materials: The roles of segregation, advection
  and diffusion, Journal of Fluid Mechanics 741 (2014) 252--279.
\newblock \href {https://doi.org/10.1017/jfm.2013.680}
  {\path{doi:10.1017/jfm.2013.680}}.

\bibitem{gray_particle_2018}
J.~M. N.~T. Gray, Particle segregation in dense granular flows, Annual Review
  of Fluid Mechanics 50 (2018) 407--433, publisher: Annual Reviews.
\newblock \href {https://doi.org/10.1146/annurev-fluid-122316-045201}
  {\path{doi:10.1146/annurev-fluid-122316-045201}}.

\bibitem{deng_modeling_2020}
Z.~Deng, Y.~Fan, J.~Theuerkauf, K.~V. Jacob, P.~B. Umbanhowar, R.~M. Lueptow,
  Modeling segregation of polydisperse granular materials in hopper discharge,
  Powder Technology 374 (2020) 389--398.
\newblock \href {https://doi.org/10.1016/j.powtec.2020.06.065}
  {\path{doi:10.1016/j.powtec.2020.06.065}}.

\bibitem{Bridgwater1969}
J.~Bridgwater, N.~W. Sharpe, D.~C. Stocker, Particle mixing by percolation,
  Transactions of the Institution of Chemical Engineers 47 (1969) T114.

\bibitem{Bridgwater1971}
J.~Bridgwater, N.~D. Ingram, Rate of spontaneous inter-particle percolation,
  Trans. Inst. Chem. Eng. 49, 163 (1971).

\bibitem{gaoPercolationFineParticle2023a}
S.~Gao, J.~M. Ottino, P.~B. Umbanhowar, R.~M. Lueptow, Percolation of a fine
  particle in static granular beds, Physical Review E 107~(1) (2023) 014903.
\newblock \href {https://doi.org/10.1103/PhysRevE.107.014903}
  {\path{doi:10.1103/PhysRevE.107.014903}}.

\bibitem{vyasImpactsPackedBed}
D.~R. Vyas, S.~Gao, P.~B. Umbanhowar, J.~M. Ottino, R.~M. Lueptow, Impacts of
  packed bed polydispersity and deformation on fine particle transport, AIChE
  Journal (2024) e18499.\href {https://doi.org/10.1002/aic.18499}
  {\path{doi:10.1002/aic.18499}}.

\bibitem{Rahman2008}
M.~Rahman, H.~Zhu, A.~Yu, J.~Bridgwater, {DEM} simulation of particle
  percolation in a packed bed, Particuology 6~(6) (2008) 475--482.
\newblock \href {https://doi.org/10.1016/j.partic.2008.07.016}
  {\path{doi:10.1016/j.partic.2008.07.016}}.

\bibitem{Li2010}
J.~Li, A.~B. Yu, J.~Bridgwater, S.~L. Rough, Spontaneous inter-particle
  percolation: A kinematic simulation study, Powder Technology 203~(2) (2010)
  397--403.
\newblock \href {https://doi.org/10.1016/j.powtec.2010.05.037}
  {\path{doi:10.1016/j.powtec.2010.05.037}}.

\bibitem{roozbahaniMechanicalTrappingFine2014a}
M.~M. Roozbahani, L.~{Graham-Brady}, J.~D. Frost, Mechanical trapping of fine
  particles in a medium of mono-sized randomly packed spheres, International
  Journal for Numerical and Analytical Methods in Geomechanics 38~(17) (2014)
  1776--1791.
\newblock \href {https://doi.org/10.1002/nag.2276}
  {\path{doi:10.1002/nag.2276}}.

\bibitem{lomineTransportSmallParticles2006}
F.~Lomin{\'e}, L.~Oger, Transport of small particles through a {{3D}} packing
  of spheres: Experimental and numerical approaches, Journal of Statistical
  Mechanics: Theory and Experiment 2006~(07) (2006) 7019.
\newblock \href {https://doi.org/10.1088/1742-5468/2006/07/P07019}
  {\path{doi:10.1088/1742-5468/2006/07/P07019}}.

\bibitem{lomineDispersionParticlesSpontaneous2009c}
F.~Lomin{\'e}, L.~Oger, Dispersion of particles by spontaneous interparticle
  percolation through unconsolidated porous media, Physical Review E 79~(5)
  (2009) 051307.
\newblock \href {https://doi.org/10.1103/PhysRevE.79.051307}
  {\path{doi:10.1103/PhysRevE.79.051307}}.

\bibitem{gaoVerticalVelocitySmall2024}
S.~Gao, J.~M. Ottino, R.~M. Lueptow, P.~B. Umbanhowar, Vertical velocity of a
  small sphere in a sheared granular bed, Physical Review Research 6~(2) (2024)
  L022015.
\newblock \href {https://doi.org/10.1103/PhysRevResearch.6.L022015}
  {\path{doi:10.1103/PhysRevResearch.6.L022015}}.

\bibitem{Drude}
P.~Drude, On the electron theory of metals, Annal. Phys 566 (1900) 306.

\bibitem{cundallDiscreteNumericalModel1979b}
P.~A. Cundall, O.~D.~L. Strack, A discrete numerical model for granular
  assemblies, G{\'e}otechnique 29~(1) (1979) 47--65.
\newblock \href {https://doi.org/10.1680/geot.1979.29.1.47}
  {\path{doi:10.1680/geot.1979.29.1.47}}.

\bibitem{osullivanParticulateDiscreteElement2014}
C.~O'Sullivan, Particulate discrete element modelling: A geomechanics
  perspective, CRC Press, London, 2014.
\newblock \href {https://doi.org/10.1201/9781482266498}
  {\path{doi:10.1201/9781482266498}}.

\bibitem{bergerChallengesIIWide2014}
K.~J. Berger, C.~M. Hrenya, Challenges of {{DEM}}: {{II}}. {{Wide}} particle
  size distributions, Powder Technology 264 (2014) 627--633.
\newblock \href {https://doi.org/10.1016/j.powtec.2014.04.096}
  {\path{doi:10.1016/j.powtec.2014.04.096}}.

\bibitem{montiLargescaleFrictionlessJamming2022}
J.~M. Monti, J.~T. Clemmer, I.~Srivastava, L.~E. Silbert, G.~S. Grest, J.~B.
  Lechman, Large-scale frictionless jamming with power-law particle size
  distributions, Physical Review E 106~(3) (2022) 034901.
\newblock \href {https://doi.org/10.1103/PhysRevE.106.034901}
  {\path{doi:10.1103/PhysRevE.106.034901}}.

\bibitem{montiFractalDimensionsJammed2023}
J.~M. Monti, I.~Srivastava, L.~E. Silbert, J.~B. Lechman, G.~S. Grest, Fractal
  dimensions of jammed packings with power-law particle size distributions in
  two and three dimensions, Physical Review E 108~(4) (2023) L042902.
\newblock \href {https://doi.org/10.1103/PhysRevE.108.L042902}
  {\path{doi:10.1103/PhysRevE.108.L042902}}.

\bibitem{ogarkoFastMultilevelAlgorithm2012}
V.~Ogarko, S.~Luding, A fast multilevel algorithm for contact detection of
  arbitrarily polydisperse objects, Computer Physics Communications 183~(4)
  (2012) 931--936.
\newblock \href {https://doi.org/10.1016/j.cpc.2011.12.019}
  {\path{doi:10.1016/j.cpc.2011.12.019}}.

\bibitem{vyasImprovedVelocityVerletAlgorithm2024}
D.~R. Vyas, J.~Ottino, R.~M. Lueptow, P.~Umbanhowar, Improved velocity-verlet
  algorithm for the discrete element method, Comput. Phys. Commun. 310 (2024)
  109524.
\newblock \href {https://doi.org/10.1016/j.cpc.2025.109524}
  {\path{doi:10.1016/j.cpc.2025.109524}}.

\bibitem{thompsonLAMMPSFlexibleSimulation2022}
A.~P. Thompson, H.~M. Aktulga, R.~Berger, D.~S. Bolintineanu, W.~M. Brown,
  P.~S. Crozier, P.~J. {in 't Veld}, A.~Kohlmeyer, S.~G. Moore, T.~D. Nguyen,
  R.~Shan, M.~J. Stevens, J.~Tranchida, C.~Trott, S.~J. Plimpton, {{LAMMPS}} -
  a flexible simulation tool for particle-based materials modeling at the
  atomic, meso, and continuum scales, Computer Physics Communications 271
  (2022) 108171.
\newblock \href {https://doi.org/10.1016/j.cpc.2021.108171}
  {\path{doi:10.1016/j.cpc.2021.108171}}.

\bibitem{thorntonInvestigationComparativeBehaviour2011c}
C.~Thornton, S.~J. Cummins, P.~W. Cleary, An investigation of the comparative
  behaviour of alternative contact force models during elastic collisions,
  Powder Technology 210~(3) (2011) 189--197.
\newblock \href {https://doi.org/10.1016/j.powtec.2011.01.013}
  {\path{doi:10.1016/j.powtec.2011.01.013}}.

\bibitem{Mindlin1949ComplianceOE}
R.~D. Mindlin, Compliance of elastic bodies in contact, Journal of Applied
  Mechanics-Transactions of the ASME 16 (1949) 259--268.
\newblock \href {https://doi.org/10.1115/1.4009973}
  {\path{doi:10.1115/1.4009973}}.

\bibitem{MARSHALL20091541}
J.~Marshall, Discrete-element modeling of particulate aerosol flows, Journal of
  Computational Physics 228~(5) (2009) 1541--1561.
\newblock \href {https://doi.org/10.1016/j.jcp.2008.10.035}
  {\path{doi:10.1016/j.jcp.2008.10.035}}.

\bibitem{shireSimulationsPolydisperseMedia2021}
T.~Shire, K.~J. Hanley, K.~Stratford, {{DEM}} simulations of polydisperse
  media: Efficient contact detection applied to investigate the quasi-static
  limit, Computational Particle Mechanics 8~(4) (2021) 653--663.
\newblock \href {https://doi.org/10.1007/s40571-020-00361-2}
  {\path{doi:10.1007/s40571-020-00361-2}}.

\bibitem{lubachevskyGeometricPropertiesRandom1990a}
B.~D. Lubachevsky, F.~H. Stillinger, Geometric properties of random disk
  packings, Journal of Statistical Physics 60~(5) (1990) 561--583.
\newblock \href {https://doi.org/10.1007/BF01025983}
  {\path{doi:10.1007/BF01025983}}.

\bibitem{Schneider1978MoleculardynamicsSO}
T.~Schneider, E.~Stoll, Molecular-dynamics study of a three-dimensional
  one-component model for distortive phase transitions, Physical Review B 17
  (1978) 1302--1322.
\newblock \href {https://doi.org/10.1103/physrevb.17.1302}
  {\path{doi:10.1103/physrevb.17.1302}}.

\end{thebibliography}
\end{document}